\documentclass[12pt]{article}
\usepackage[T1]{fontenc}
\usepackage{amsmath}
\usepackage[section]{placeins}
\usepackage{rotating}
\usepackage{color}
\usepackage[english]{babel}
\usepackage{standalone}
\usepackage{float}
\usepackage{booktabs}
\usepackage{amsmath}
\usepackage{rotating}
\usepackage{graphics}
\graphicspath{ {image/} }
\usepackage{listings}
\usepackage{setspace}
\usepackage[super,numbers,sort&compress]{natbib}
\bibpunct{}{}{,}{a}{,}{,}
\setcitestyle{numbers}
\usepackage{url}
\usepackage[titletoc,title]{appendix}
\usepackage[left=1.5cm,right=1.5cm,top=2cm,bottom=2cm]{geometry}
\usepackage[titletoc]{appendix}
\usepackage{tikz}
 
\usetikzlibrary{fit,positioning}
\usetikzlibrary{shapes}
\usetikzlibrary{arrows, decorations.markings, matrix}
\usetikzlibrary{fit}
\usetikzlibrary{chains}
\usetikzlibrary{arrows}
\usepackage[font=footnotesize, labelfont={bf}, margin=1cm]{caption}
\usetikzlibrary {trees}
\usetikzlibrary{arrows,positioning,shapes.geometric}
\usepackage{tikz-qtree}
\usepackage{rotating}
\usepackage{array}
\usepackage{colortbl}
\usepackage{enumerate,multicol}
\usepackage{amsmath, amssymb, amsthm}
\usepackage{enumitem,bm}
\usepackage{parcolumns}

\providecommand{\keywords}[1]{{\bf{Keywords:}} #1}

\usepackage[multiple]{footmisc}
\usepackage{authblk}
\usepackage{etoolbox}
\usepackage[position=bottom]{subfig}
\usepackage{multirow}
\usepackage{makecell}
\usepackage{fancyvrb} 
\usepackage{listings,color} 
\usepackage{moreverb}

\newcommand\ack{\section*{Acknowledgements}}

\addto{\captionsenglish}{}

\usepackage{pifont}
\newcommand{\cmark}{\ding{51}}%
\newcommand{\xmark}{\ding{55}}%

\begin{document}

\title{\bf {A Bayesian framework for patient-level partitioned survival cost-utility analysis}}
\author[1]{Andrea Gabrio\thanks{E-mail: andrea.gabrio.15@ucl.ac.uk}}
\affil[1]{\small \textit{Department of Statistical Science, University College London,~UK}}

\date{}
\maketitle

\vspace*{-1cm}
\abstract{Patient-level health economic data collected alongside clinical trials are an important component of the process of technology appraisal, with a view to informing resource allocation decisions. For end of life treatments, such as cancer treatments, modelling of cost-effectiveness/utility data may involve some form of partitioned survival analysis, where measures of health-related quality of life and survival time for both pre- and post-progression periods are combined to generate some aggregate measure of clinical benefits (e.g.~quality-adjusted survival). In addition, resource use data are often collected from health records on different services from which different cost components are obtained (e.g.~treatment, hospital or adverse events costs). A critical problem in these analyses is that both effectiveness and cost data present some complexities, including non-normality, spikes, and missingness, that should be addressed using appropriate methods. Bayesian modelling provides a powerful tool which has become more and more popular in the recent health economics and statistical literature to jointly handle these issues in a relatively easy way. This paper presents a general Bayesian framework that takes into account the complex relationships of trial-based partitioned survival cost-utility data, potentially providing a more adequate evidence for policymakers to inform the decision-making process. Our approach is motivated by, and applied to, a working example based on data from a trial assessing the cost-effectiveness of a new treatment for patients with advanced non-small-cell lung cancer.}

\vspace*{0.5cm}
\vspace*{0.5cm}
\hrule

\section{Introduction}\label{intro}
Constraints on healthcare resource, particularly during times of economic turmoil and instability, result in governments taking a hard look at all public expenditure, including the medicines budget. As a result, in many countries, government bodies who are responsible for the assessment of the costs relative to the effects of specific treatments now require evidence of economic value as part of their reimbursement decision-making. For example, in the United Kingdom (England and Wales), this assessment is made by the National Institute for Clinical Health Excellence (NICE)~\textsuperscript{\citep{earnshaw2008nice}}. The statistical analysis of health economic data has become an increasingly important component of clinical trials, which provide one of the earliest opportunities to generate economic data that can be used for~decision-making~\textsuperscript{\citep{glick2014economic}}. 

The standard analysis of individual-level data involves the comparison of two interventions for which suitable measures of effectiveness and costs are collected on each patient enrolled in the trial, often at different time points throughout the follow-up period. Depending on the economic perspective of the analysis, different types of resource use data (e.g.~hospital visits, consultations, scans, number of doses, etc.) are collected for each patient and time point using electronic health records, self-reported questionnaires or a combination of both. Service use information is combined with unit prices taken from national sources to derive patient-level costs associated with each type of service and the total costs over the trial duration are obtained by summing up all cost components at each time. The effectiveness is often measured in terms of preference-based health-related quality of life instruments (e.g.~the EQ-5D questionnaires~\textsuperscript{\citep{EQ5D}}) and are combined with national tariff systems to value the patients' health states in terms of \textit{utility scores}, measured on a scale from $0$ (worst imaginable health) to $1$ (perfect health), using the preferences of a sample of the public~\textsuperscript{\citep{dolan1997modeling}}. States considered worse than death are also possible and are associated with values below $0$. Finally, a single metric, called \textit{Quality-Adjusted Life Years} (QALYs), is calculated by aggregating the utility scores over the trial duration and represents the health outcome of choice in the economic analysis. A common approach for calculating a QALY is to compute the \textit{Area Under the Curve} (AUC)~\citep{Drummond}:. 
\begin{equation}\label{QALY}
\text{QALY}_{it}=\sum_{j=1}^{J}\left(\frac{u_{ijt}+u_{ij-1t}}{2}\right)\delta_{j},
\end{equation}
where $u_{ijt}$ is the utility score for the $i$-th patient in treatment $t$ at the $j$-th time in the trial, while $\delta_{j}=(\text{Time}_{j}-\text{Time}_{j-1})/ (\text{Unit of time})$ is the fraction of the time unit (typically $1$ year) which is covered between time $j-1$ and $j$. For example, if measurements in the trial are collected at monthly intervals, then $\delta_j=(\text{1 month})/ (\text{12 months})=0.083$. We note that for patients who die at some point in the trial, which is often the case for cancer trials, assumptions have to be made about their utility values at all time points after the time of death so that their QALYs can be computed using Equation~\ref{QALY}. In many cases, a utility of $0$ is typically associated with a state of death at a given time point and is carried over until the last follow-up, which may cause an underestimation of the AUC. However, this assumption may not be unreasonable, because often disease progression precedes death, and the health state is expected to be worse (even worse than death) around this~time~\textsuperscript{\citep{khan2015design}}.

The AUC method is a relatively simple approach but assumes an absence of missing values/censoring. When the primary end point of the trial is survival, such as in cancer trials where patients may be either dead or still alive at the end of the study (often censored), a possible way to deal with censoring is to combine the information from both utility and survival for each patient. More specifically, it is possible to multiply each patient's observed survival at time $j$ by his/her corresponding utility values at the same time to formulate a QALY end point on an AUC scale, also known as \textit{Quality Adjusted Survival} (QAS), that is
\begin{equation}\label{QALY2}
\text{QAS}_{it}=\sum_{j=1}^{J}\left(\frac{{u}_{ijt}+{u}_{ij-1t}}{2}\right)\left(\frac{s_{ijt}+s_{ij-1t}}{2}\right)\delta_{j},
\end{equation}
where ${u}_{ijt}$ and $s_{ijt}$ are the utility and the survival time for the $i$-th patient in treatment $t$ at the $j$-th time. Equation~\ref{QALY2} can be thought as a time to event analysis using the QALY as the analysis end point, given that it is still in a time measurement, albeit weighted~\textsuperscript{\citep{glasziou1990quality}}. When some utility scores are missing (e.g.~due to incomplete questionnaire), the empirical values ${u}_{ijt}$ can be replaced with predicted individual-level utilities which are generated from some model. This allows to extrapolate the utility scores for those patients still alive in the future, predict the values at different time points within the follow-up period, and adjust for possible imbalances between treatments in some baseline variables. However, care should be taken that predictions are computed sensibly: if a patient died at time $j$, then his/her predicted utility estimates from at $j$ and all subsequent times should be equal to~zero.

\subsection{Partitioned Survival Cost-Utility Analysis}\label{ps_cua}
One important aspect, particularly in cancer trials, is that survival time can change rapidly after the disease has progressed. Therefore, inferences about mean utilities should take into account the specific features of pre- and post-progression responses as well as their dependence relationships. This is the rationale behind \textit{partitioned survival analysis}, which involves the partitioning of survival data for the time to event end point, typically~\textit{Overall Survival} (OS), into two components: \textit{Progression Free Survival} (PFS) and \textit{Post-Progression Survival} (PPS), where $\text{OS} = \text{PFS} + \text{PPS}$. Hence, QAS based on the AUC approach can alternatively be computed during the pre- and post-progression periods by multiplying each survival component by the pre- and post-progression utilities. The partitioning of health-related quality of life data based on different components of survival time forms the basis for what is known as \textit{Partitioned Survival Cost-Utility Analysis}, where patient-level QAS based on OS data ($\text{QAS}_{it}^{\text{OS}}$) can be expressed as
\begin{equation}\label{QALY3}
\text{QAS}^{\text{OS}}_{it}=\text{QAS}^{\text{PFS}}_{it} + \text{QAS}^{\text{PPS}}_{it},
\end{equation}
where $\text{QAS}^{\text{PFS}}_{it}$ and $\text{QAS}^{\text{PPS}}_{it}$ are the QAS computed as in Equation~\ref{QALY2} using patient-level utilities and survival times for pre- and post-progression periods, respectively. In most analyses, the different survival components in Equation~\ref{QALY3} are separately modelled using parametric regression models to allow for extrapolation~\textsuperscript{\citep{khan2015design,gelber1995comparing}}. However, in many cancer trials, direct modelling of $\text{QAS}^{\text{PPS}}_{it}$ is not possible since utility data are collected only up to progression, while utilities after disease progressions are extrapolated based on some modelling assumptions and OS or PFS data~\textsuperscript{\citep{williams2017estimation}}. However, if PPS is collected, it is generally recommended to use whatever of these data are available to improve the estimate of $\text{QAS}^{\text{PPS}}_{it}$. 

It is important to note that the calculation of QAS in Equation~\ref{QALY2} and Equation~\ref{QALY3} assumes an absence of censoring. In practice, however, some of the patients may be still alive at the end of the trial (censored), which may hinder the validity of standard approaches for calculating QAS data. Indeed, on a QALY scale, survival times may be altered (by the utility weights), resulting in \textit{informative censoring} (i.e.~related to survival time), which makes the standard Kaplan-Meier analysis invalid~\textsuperscript{\citep{glasziou1990quality}}. For the rest of the paper and in the analysis of our case study, we will assume that no informative censoring occurs (in the case study $> 99\%$ of patients had died at the time of the analysis) so that standard partitioned survival cost-utility analysis methods for computing QAS data can be assumed to be valid. In Section~\ref{discussion} we will discuss the potential implications associated with informative censoring on these types of analyses as well as some alternative approaches that can be used to overcome the limitations associated with standard approaches in this context. 

\subsection{Modelling of Patient-Level Cost-Utility Data}\label{model_qas}
Statistical modelling for trial-based cost-effectiveness/utility data has received much attention in both the health economics and the statistical literature in recent years~\textsuperscript{\citep{willan2006statistical,ramsey2015cost}}, increasingly often under a Bayesian statistical approach~\textsuperscript{\citep{o2001bayesian,spiegelhalter2004bayesian,Baioa}}. From the statistical point of view, this is a challenging problem because of the generally complex relationships linking the measure of effectiveness (e.g.~QALYs) and the associated costs. First, the presence of a bivariate outcome requires the use of appropriate methods taking into account correlation~\textsuperscript{\citep{Nixon,Grieveb,Gomes2011}}. Second, both utility and cost data are characterised by empirical distributions that are highly skewed and simplifying assumptions, such as (bivariate) normality of the underlying distributions, are usually not granted. The adoption of parametric distributions that can account for skewness (e.g.~beta for the utilities and gamma or log-normal distributions for the costs) has been suggested to improve the fit of the model~\textsuperscript{\citep{OHagan,Thompson,Basu}}. Third, data may exhibit spikes at one or both of the boundaries of the range for the underlying distribution, e.g.~null costs and perfect health (i.e.~utility of one), that are difficult to capture with standard parametric models~\textsuperscript{\citep{mihaylova2011review,Basu}}. The use of more flexible formulations, known as \textit{hurdle models}, has been recommended to explicitly account for these "structural" values~\textsuperscript{\citep{baio2014bayesian,gabrio2019full,gabrio2019bayesian}}. These models consist in a mixture between a point mass distribution (the spike) and a parametric model fit to the natural range of the relevant variable without the boundary values. 
Finally, individual-level data from clinical trials are almost invariably affected by the problem of missing data. Analyses that are limited
to individuals with fully-observed data (complete case analysis) are inefficient and assume that the completers are a random sample of all individuals, an assumption known as \textit{Missing Completely At Random} (MCAR), which may yield biased inferences~\textsuperscript{\citep{Rubina}}. Alternative and more efficient approaches, such as multiple imputation and likelihood-based methods~\textsuperscript{\citep{Rubina, molenberghs2014handbook}}, rely on the less restrictive assumption that all observed data can be used to explain fully the reason for why some observations are missing, an assumption known as Missing At Random (MAR). However, the possibility that unobserved data influence the mechanism by which the missing values arise, an assumption known as  \textit{Missing Not At Random} (MNAR), can never be ruled out and will introduce bias when inferences are based on the observed data alone. Content-specific knowledge and tailored modelling approaches can be used to make inferences under MNAR and, within a Bayesian approach, informative prior distributions represent a powerful tool for conducting sensitivity analysis to different missingness assumptions~\textsuperscript{\citep{Daniels}}.


In this paper, we aim at extending the current methods for modelling trial-based partitioned survival cost-utility data, taking advantage of the flexibility of Bayesian methods, and specify a joint probabilistic model for the health economic outcomes. We propose a general framework that is able to account for the multiple types of complexities affecting individual level data (correlation, missingness, skewness and structural values), while also explicitly modelling the dependence relationships between different types of quality of life and cost components. The paper is structured as follows: first, in Section~\ref{methods}, we set out our modelling framework. In Section~\ref{topical} we present the case study used as motivating example and describe the data, while Section~\ref{application} defines the general structure of the model used in the analysis and how it is tailored to deal
with the different characteristics of the data. Section~\ref{results} presents and discusses the statistical and health economic results of the analysis. Finally, Section~\ref{discussion} summarises our conclusions.

\section{Methods}\label{methods}
Consider a clinical trial in which patient-level information on a set of suitably defined effectiveness and cost variables has been collected at $J$ time points on $N$ individuals, who have been allocated to $T$ intervention groups. We assume that the primary endpoint of the trial is OS, while secondary endpoints include PFS, a self-reported health-related quality of life questionnaire (e.g.~EQ-5D) and health records on different types of services (e.g.~drug frequency and dosage, hospital visits, etc.) for each individual. We denote with $\bm e_{it}$ the set of effectiveness variables for the $i$-th individual in treatment $t$ of the trial, which are obtained by combining survival time and utility scores (e.g.~QAS). We assume that utility data are collected up to and beyond progression so that both $\text{QAS}^{\text{PFS}}$ and $\text{QAS}^{\text{PPS}}$ can be obtained from the observed data without extrapolation. We denote the full set of effectiveness variables as $\bm e_{it}=(e^{\text{PFS}}_{it},e^{\text{PPS}}_{it})$, formed by the pre- and post-progression components. Next, we denote with $\bm c_{it}$ the set of patient-level cost variables, obtained by combining the resource use data on $K$ different services over the duration of the trial with a unit prices for each type of service. The full set of cost variables can therefore be expressed as $\bm c_{it}=(c^1_{it},\ldots,c^K_{it})$. Finally, it is also common to have some patient-level information on a set of additional variables $\bm x_{it}$ (for example on age, sex or potential co-morbidities) which may be included in the economic analysis. Without loss of generality, we assume in the following that only two interventions are compared: $t=1$ is some standard (e.g.~currently recommended or applied by the health care provider), and $t=2$ is a new intervention being suggested to potentially replace the~standard. 

The objective of the economic evaluations is to perform a patient-level partitioned survival cost-utility analysis by specifying a joint model $p(\bm e_{it}, \bm c_{it} \mid \bm \theta)$, indexed by a set of parameters $\bm \theta$ comprising the marginal mean effectiveness and cost parameters $\bm \mu=(\mu_{et},\mu_{ct})$ on which inference has to be made to inform the decision-making process. Different approaches can be used to specify $p(\bm e_{it}, \bm c_{it})$. Here, we take advantage of the flexibility granted by the Bayesian framework and express the joint distribution as 
\begin{equation}\label{factor_ec}
p(\bm e_{it}, \bm c_{it}) = p(\bm e_{it})p(\bm c_{it} \mid \bm e_{it}),
\end{equation}
where $p(\bm e_{it})$ is the \textit{marginal distribution} of the effectiveness and $p(\bm c_{it} \mid \bm e_{it})$ is the \textit{conditional distribution} of the costs given the effectiveness. Note that it is always possible to specify the joint distribution using the reverse factorisation $p(\bm e_{it}, \bm c_{it}) = p(\bm c_{it})p(\bm e_{it} \mid \bm c_{it})$ but, following previous works~\textsuperscript{\citep{lambert2008estimating,gabrio2019full}}, we describe our approach through a marginal distribution for the effectiveness and a conditional distribution for the costs. In the following, we describe how the two factors on the right-hand side of Equation~\ref{factor_ec} are~specified.
 
\subsection{Marginal Model for the Effectiveness}\label{marginal_e}
For each individual and treatment, we specify a marginal distribution of the effectiveness variables $\bm e_{it}=(e^{\text{PFS}}_{it},e^{\text{PPS}}_{it})$ using the conditional factorisation:
\begin{equation}\label{factor_e}
p(\bm e_{it} \mid \bm \theta_{et} )=p(e^{\text{PFS}}_{it} \mid \bm \theta^{\text{PFS}}_{et})p(e^{\text{PPS}}_{it} \mid e^{\text{PFS}}_{it}, \bm \theta^{\text{PPS}}_{et}),
\end{equation}
where $\bm \theta_{et}=(\bm \theta^{\text{PFS}}_{et}, \bm \theta^{\text{PPS}}_{et})$ are the treatment-specific effectiveness parameters formed by the two distinct sets that index the marginal distribution of $e^{\text{PFS}}_{it}$ and the conditional distribution of $e^{\text{PPS}}_{it} \mid e^{\text{PFS}}_{it}$. The sets of parameters $\bm \theta_{et}$ can be expressed in terms of some \textit{location} $\bm \phi_{iet}=(\phi^{\text{PFS}}_{iet},\phi^{\text{PPS}}_{iet})$ and \textit{ancillary} $\bm \psi_{et}=(\bm \psi^{\text{PFS}}_{et},\bm \psi^{\text{PPS}}_{et})$ parameters, the latter typically comprising some standard deviations $\bm \sigma_{et}=(\sigma^{\text{PFS}}_{et},\sigma^{\text{PPS}}_{et})$. Often, interest is in the modelling of the location parameters as a function of other variables. This is typically achieved through a generalised linear structure and some link function that relates the expected value of the response to the linear predictors in the model. For example, we can use some parametric model to specify the distribution of the two effectiveness variables as
 \begin{equation}\label{dist_e}
e^{\text{PFS}}_{it} \sim f^{\text{PFS}}(\phi^{\text{PFS}}_{iet},\bm \psi^{\text{PFS}}_{et}) \;\;\; \text{and} \;\;\; e^{\text{PPS}}_{it} \mid e^{\text{PFS}}_{it}  \sim f^{\text{PPS}}(\phi^{\text{PPS}}_{iet},\bm \psi^{\text{PPS}}_{et}),
\end{equation}
where $f^{\text{PFS}}(\cdot)$ and $f^{\text{PPS}}(\cdot)$ denote some generic parametric distribution (e.g.~normal) used to model $e^{\text{PFS}}_{it}$ and $e^{\text{PFS}}_{it}\mid e^{\text{PFS}}_{it}$, respectively. The locations of the two models in Equation~\ref{dist_e} are then modelled~as:
 \begin{equation}\label{glm_mu_e}
\begin{split}
g(\phi^{\text{PFS}}_{iet})&=\alpha^{\text{PFS}}_{0t} + [ \ldots ],\\[2ex]
g(\phi^{\text{PPS}}_{iet})&=\alpha^{\text{PPS}}_{0t} + \alpha^{\text{PPS}}_{1t}(e^{\text{PFS}}_{it} - \mu^{\text{PFS}}_{et}) + [ \ldots ],
\end{split}
\end{equation}
where $g(\cdot)$ is the link function, $\bm  \alpha^{\text{PFS}}= (\alpha^{\text{PFS}}_{0t}, \ldots)$ and $\bm  \alpha^{\text{PPS}}= (\alpha^{\text{PPS}}_{0t}, \alpha^{\text{PPS}}_{1t}, \ldots)$ are the sets of regression parameters indexing the two models, while the notation $+ \; [\ldots]$ indicates that other terms (e.g.~quantifying the effect of relevant covariates $\bm x_{it}$) may or may not be included in each model. In the absence of covariates or assuming that a centred version $(\bm x_{it}-\bar{\bm x_t})$ is used, the quantities $\mu^{\text{PFS}}_{et}=g^{-1}(\alpha^{\text{PFS}}_{0})$ and $\mu^{\text{PPS}}_{et}=g^{-1}(\alpha^{\text{PPS}}_{0})$ can be interpreted as the population mean effectiveness for  $e^{\text{PFS}}$ and $e^{\text{PPS}}$, respectively.

\subsection{Conditional Model for the Costs Given the Effectiveness}\label{conditional_ce}
For the conditional distribution of the costs given the effectiveness, we use a similar approach to the one of $\bm e_{it}$ and factor $\bm c_{it} \mid \bm e_{it}$ as the product of a sequence of $K$ conditional~distributions: 
\begin{equation}\label{factor_ce}
p(\bm c_{it} \mid \bm e_{it}, \bm \theta_{ct} )=p(c^{1}_{it} \mid \bm e_{it}, \bm \theta^{1}_{ct}) \cdot  \cdot \cdot  p(c^{K}_{it} \mid \bm e_{it}, c^1_{it}, \ldots, c^{K-1}_{it}, \bm \theta^{K}_{ct}),
\end{equation}
where $\bm \theta_{ct}=(\bm \theta^{1}_{ct},\ldots, \bm \theta^{K}_{ct})$ are the treatment-specific cost parameters that index the $K$ conditional distributions of the cost variables, with $K$ being the number of cost components. These parameters can also be expressed in terms of a series of $K$ location $\bm \phi_{ict}=(\phi^1_{ict},\ldots, \phi^K_{ict})$ and ancillary $\bm \psi_{ct}=(\psi^1_{ct},\ldots,\psi^K_{ct})$ parameters, the latter typically including a vector of standard deviations $\bm \sigma_{ct}=(\sigma^1_{ct},\ldots,\sigma^K_{ct})$. We can model the univariate distributions of the cost variables using some parametric forms: 
 \begin{equation}\label{dist_ce}
c^{1}_{it} \mid \bm e_{it} \sim f^{1}(\phi^{1}_{ict},\bm \psi^{1}_{ct}), \;\;\;  \cdot  \cdot \cdot \, ,   \;\;\; c^{K}_{it} \mid  \bm e_{it}, c^{1}_{it},\ldots, c^{K-1}_{it}  \sim f^{K}(\phi^{K}_{ict},\bm \psi^{K}_{ct}),
\end{equation}
where $f^{1}(\cdot), \ldots,f^{K}(\cdot)$ denote the parametric distributions used to model the patient-level cost data for the $K$ cost components. The location of each variable can then be modelled as a function of other variables using the generalised linear forms:
 \begin{equation}\label{glm_mu_ce}
\begin{split}
g(\phi^{1}_{ict})&=\beta^{1}_{0t} + \beta^{1}_{1t}(e^{\text{PFS}}_{it} - \mu^{\text{PFS}}_{et}) +\beta^{1}_{2t}(e^{\text{PPS}}_{it} - \mu^{\text{PPS}}_{et})  + [ \ldots ],\\[2ex]
 & \Large \vdots\\[2ex]
g(\phi^{K}_{ict})&=\beta^{K}_{0t} + \beta^{K}_{1t}(e^{\text{PFS}}_{it} - \mu^{\text{PFS}}_{et}) +\beta^{K}_{2t}(e^{\text{PPS}}_{it} - \mu^{\text{PPS}}_{et}) \; + \\[2ex]
& \quad \; \beta^{K}_{3t}(c^{1}_{it} - \mu^{1}_{ct}) + \ldots + \beta^{K}_{K+1,t}(c^{K-1}_{it} - \mu^{K-1}_{ct}) + [ \ldots ],
\end{split}
\end{equation}
where $\bm  \beta^{1}= (\beta^{1}_{0t}, \beta^{1}_{1t}, \beta^{1}_{2t}, \ldots), \ldots, \bm  \beta^{K}= (\beta^{K}_{0t}, \beta^{K}_{1t}, \beta^{K}_{2t}, \beta^{K}_{3t}, \ldots, \beta^{K}_{K+1,t} \ldots)$ are the sets of regression parameters indexing the $K$ models. Assuming other covariates are also either centred or absent, the quantities $\mu^{1}_{ct}=g^{-1}(\beta^{1}_{0t}), \ldots, \mu^{K}_{ct}=g^{-1}(\beta^{K}_{0t})$ in Equation~\ref{glm_mu_ce} can be interpreted as the $K$ population mean cost components. 

Figure~\ref{model_frame} provides a visual representation of the general modelling framework described above.
\begin{figure}[!h]
\centering
FIGURE 1
\label{model_frame}
\end{figure}
The effectiveness and cost distributions are represented in terms of combined "modules"- the red and blue boxes - in which the random quantities are linked through logical relationships. The links both within and between the variables in the modules ensure the full characterisation of the uncertainty in the model. Notably, this is general enough to be extended to any suitable distributional assumption, as well as to handle covariates in either or both the modules. 
In the following section, we first introduce and describe the case study that we use as motivating example. Next, we present the modelling specification for the effectiveness and cost data chosen in our analysis. The model is specified so to take into account different issues of the individual-level data, including correlation between variables, missingness, skewness and presence of structural~values. 

\section{Example: The TOPICAL trial}\label{topical}
The TOPICAL study is a double-blind, randomised, placebo-controlled, phase III trial, conducted in the UK in predominantly elderly patients with non-small-cell lung cancer receiving best supportive care considered unfit for chemotherapy because of poor performance status and/or multiple medical comorbidities~\textsuperscript{\citep{lee2012first}}. Subjects were randomly assigned to receive a control (oral placebo) ($t=1$) or erlotinib ($150$ mg per day, $t=2$) until disease progression or unacceptable toxicity. All patients were allowed to receive immediate or delayed palliative chest radiotherapy and/or radiotherapy to metastatic sites as appropriate. The original trial investigated $350$ patients in the active treatment and $320$ in the placebo group, with time of interest for the cost-effectiveness analysis being $1$ year. Patient-level QAS, cost are measured and available to us, for a subsample of $300$ patients ($150$ in the placebo and $150$ in the erlotinib group, respectively). 

The primary end point of the trial was OS; secondary end points were PFS (defined as the time between randomisation and progression or death), and health-related quality of life measured by the EQ-5D-3L questionnaire which was collected at monthly intervals. Patient-level QAS for both PFS ($e^{\text{PFS}}_{it}$) and PPS ($e^{\text{PPS}}_{it}$) were determined using the EQ-5D utilities over time for pre- and post-progression periods and multiplied by corresponding survival times. Since all patients in our dataset progressed/died by the time of the analysis, no extrapolation of OS and PFS was carried out but utility data were obtained prospectively up to and beyond progression~\textsuperscript{\citep{khan2015cost}}. The costs are from three main components: 1) drug (erlotinib), radiotherapy, additional anticancer treatments - denoted with $c^{\text{drug}}_{it}$; 2) patient management (e.g.~hospital visits) - denoted with $c^{\text{hos}}_{it}$; 3) and managing important treatment related-adverse events (e.g.~rash) - denoted with $c^{\text{ae}}_{it}$. Resource use was collected monthly on the case report forms and combined with unit prices from published sources to derive the costs for each component. Figure~\ref{hist_e} and Figure~\ref{hist_c} show the histograms of the distributions of the different components of the observed QAS and cost data in both treatment groups,~respectively. The number of observations, the empirical mean and standard deviations for each variable are reported in the graphs. 
\begin{figure}[!h]
\centering
FIGURE 2
\end{figure}
The observed distributions of $e^{\text{PFS}}_{it}$ and $e^{\text{PPS}}_{it}$ show a considerable degree of skewness in both treatment groups, especially for post-progression QAS data. Although $\bm e$ are defined on a similar range of values, $e^{\text{PFS}}_{it}$ can take both negative and positive values, while $e^{\text{PPS}}_{it}$ has a lower bound at zero, with about $50\%$ of the individuals in each group being associated with this boundary (structural value).
\begin{figure}[!h]
\centering
FIGIRE 3
\end{figure}
The observed distributions of $c^{\text{drug}}_{it}$, $c^{\text{hos}}_{it}$ and $c^{\text{ae}}_{it}$ show a high degree of skewness, especially in the intervention group. All $\bm c$ data are defined on a positive range but each component has a different variability, with $c^{\text{drug}}$ in the intervention being the component associated with the largest standard deviation. The proportions of individuals who are associated with a structural zero for each cost component are: $60\%$ (only in the control group) for $c^{\text{drug}}_{it}$, $25\%$ (in each group) for $c^{\text{hos}}_{it}$, and $18\%$ (in each group) for $c^{\text{ae}}_{it}$. The total number of individuals with fully-observed data for all variables (completers) is $249 (83\%)$, while among those with partially-observed data ($51 (27\%)$) the majority were associated with unobserved values for either $e^{\text{PFS}}_{it}$, $e^{\text{PPS}}_{it}$ or $c^{\text{drug}}_{it}$ or a combination of these ($29/51=57\%$). A detailed presentation of the missingness patterns and numbers of individuals in each pattern for these data is reported in Appendix~\ref{A1}. We note that missingness in $\bm e$ is only due to incomplete EQ-5D questionnaires (and thus utility scores) and not censoring of survival time, since all patients progressed/died by the time of the analysis. Missingness in $\bm c$ is due to incomplete information from the case report forms on resource use. 



\section{Application to the TOPICAL study}\label{application}

\subsection{Model Specification}\label{model_spec}
Throughout, we refer to our motivating example to demonstrate the flexibility of our approach in dealing with the different issues affecting the cost-utility data; we note that these are likely to be encountered in many practical cases, thus making our arguments applicable in general. We start by modelling $e^{\text{PFS}}_{it}$ with a Gumbel distribution using an identify link function for the mean:
\begin{equation}\label{eq_pfs}
\begin{split}
e^{\text{PFS}}_{it} &\sim \text{Gumbel}(\phi^{\text{PFS}}_{et},\sigma^{\text{PFS}}_{et}),\\ 
\phi^{\text{PFS}}_{et} & = \alpha^{\text{PFS}}_{0t},
\end{split}
\end{equation}
where $\phi^{\text{PFS}}_{et}$ and $\sigma^{\text{PFS}}_{et}$ are the mean and standard deviation of $e^{\text{PFS}}_{it}$. We choose the Gumbel distribution since it was the parametric model associated with the best fit to the observed $e^{\text{PFS}}_{it}$ among those assessed (which included Logistic and Normal distributions). The Gumbel distribution has already been recommended for modelling utility data as it is defined on the real line while also being able to capture skewness~\textsuperscript{\citep{gomes2019copula}}.We parameterise the Gumbel distribution in terms of mean and standard deviation to facilitate the specification of the priors on the parameters, compared with using the canonical location $a$ (real) and scale $b>0$ parameters. More specifically, the mean and standard deviation of the Gumbel distribution are linked to the canonical parameters through the relationships: $a=\phi - b\kappa$ and $b=(\sigma \sqrt{6})/ \pi$, where $\kappa$ is the Euler's constant. 

When choosing the model for $e^{\text{PPS}}_{it}$ it is important to take into account the considerable proportion of people associated with a zero value in the empirical distributions of both treatment groups~(Figure~\ref{hist_e}), which may otherwise lead to biased inferences. We propose to handle this using a hurdle approach, where the distribution of $e^{\text{PPS}}_{it}$ is expressed as a mixture of a point mass distribution at zero and a parametric model for the natural range of the variable excluding the zeros. Specifically, for each subject we define an indicator variable $d^{PPS}_{it}$ taking value $1$ if the $i$-th individual is associated with $e^{\text{PPS}}_{it}=0$ and $0$ otherwise (i.e.~$e^{\text{PFS}}_{it}>0$). We then model the conditional distribution of $d^{\text{PPS}}_{it} \mid e^{\text{PFS}}_{it}$ with a Bernoulli distribution using a logit link function for the probability of being associated with a zero:
\begin{equation}\label{eq_pps_hu}
\begin{split}
d^{\text{PPS}}_{it} \mid e^{\text{PFS}}_{it}&\sim \text{Bernoulli}(\pi^{\text{PPS}}_{iet}),\\ 
\text{logit}(\pi^{\text{PPS}}_{iet}) & = \gamma^{\text{PFS}}_{0t} + \gamma^{\text{PPS}}_{1t}e^{\text{PFS}}_{it},
\end{split}
\end{equation}
where $\pi^{\text{PPS}}_{iet}$ is the probability associated with $e^{\text{PPS}}_{it}=0$, which is expressed as a linear function of $e^{\text{PFS}}_{it}$ on the logit scale via the intercept and slope regression parameters $ \gamma^{\text{PPS}}_{0t}$ and $ \gamma^{\text{PPS}}_{1t}$, respectively. Other covariates which are thought to be strongly associated with the chance of having a zero can also be included in the logistic regression to improve the estimation of the probability parameter. However, in our analysis, the inclusion of any of the baseline variables available in the trial did not lead to substantial changes in the inferences, while also not improving the fit of the model to the observed data compared with Equation~\ref{eq_pps_hu}. Therefore, we decided to remove these variables and keep the current specification for the model of $d^{\text{PPS}}_{it}$. We model $e^{\text{PPS}}_{it} \mid d^{\text{PPS}}_{it}=0, e^{\text{PFS}}_{it}$ with an Exponential distribution using a log link function for the conditional~mean:
\begin{equation}\label{eq_pps}
\begin{split}
e^{\text{PPS}}_{it} \mid d^{\text{PPS}}_{it}=0, e^{\text{PFS}}_{it}&\sim \text{Exponential}(\phi^{\text{PPS}}_{iet}),\\ 
\log(\phi^{\text{PPS}}_{iet}) & = \alpha^{\text{PPS}}_{0t} + \alpha^{\text{PPS}}_{1t}e^{\text{PFS}}_{it},
\end{split}
\end{equation}
where $ \alpha^{\text{PFS}}_{0t}$ and $\alpha^{\text{PPS}}_{1t}$ are the intercept and slope mean regression parameters for $e^{\text{PPS}}_{it}>0$, defined on the log scale. Again, the choice of the Exponential distribution was made according to the fit to the observed $e^{\text{PPS}}_{it}$ after comparing alternative model specifications (including Weibull and Normal distributions). We note that the canonical rate parameter $r$ of the Exponential distribution can be retrieved from the mean parameter through the relationship: $r=\frac{1}{\phi}$.

Using a similar modelling structure to that of $e^{\text{PPS}}_{it}$, we specify the models for the conditional distributions of the cost variables $\bm c_{it}=(c^{\text{drug}}_{it},c^{\text{hos}}_{it},c^{\text{ae}}_{it})$. We use a hurdle approach to handle the "structural" zero components for each variable, and fit Lognormal distributions to the positive costs (chosen in light of the better fit to the observed data compared with Gamma distributions). For each modelled cost variable, we checked whether the inclusion of any of the available baseline covariates from the trial could lead to some model improvement in terms of fit to the observed data or parameter estimates. However, results from the different model specifications suggest that there is no substantial gain from including these variables, which were therefore removed. We model the conditional distribution of the zero drug cost indicators and drug cost variables given $\bm e_{it}$ as:
 \begin{equation}\label{eq_drug}
\begin{split}
d^{\text{drug}}_{it} \mid \bm e_{it} &\sim \text{Bernoulli}(\pi^{\text{drug}}_{ict}),\\ 
\text{logit}(\pi^{\text{drug}}_{ict}) & = \delta^{\text{drug}}_{0t} + \delta^{\text{drug}}_{1t}e^{\text{PFS}}_{it} + \delta^{\text{drug}}_{2t}e^{\text{PPS}}_{it},\\[1ex]
c^{\text{drug}}_{it} \mid d^{\text{drug}}_{it}=0, \bm e_{it} &\sim \text{Lognormal}(\phi^{\text{drug}}_{ict}, \sigma^{\text{drug}}_{ct}),\\ 
\phi^{\text{drug}}_{ict} & = \beta^{\text{drug}}_{0t} + \beta^{\text{drug}}_{1t}e^{\text{PFS}}_{it} + \beta^{\text{drug}}_{2t}e^{\text{PPS}}_{it},
\end{split}
\end{equation}
where $\pi^{\text{drug}}_{ict}$ is the probability of having $c^{\text{drug}}_{it}=0$, while $\phi^{\text{drug}}_{ict}$ and $\sigma^{\text{drug}}_{ct}$ are the mean and standard deviation parameter for $c^{\text{drug}}_{it}>0$ on the log scale. The regression parameters $\bm \delta^{\text{drug}} = (\delta^{\text{drug}}_{0t}, \delta^{\text{drug}}_{1t}, \delta^{\text{drug}}_{2t})$ and $\bm \beta^{\text{drug}} = (\beta^{\text{drug}}_{0t}, \beta^{\text{drug}}_{1t}, \beta^{\text{drug}}_{2t})$ capture the dependence between drug costs and the effectiveness variables for the zero and non-zero components, respectively. The conditional distribution of the zero hospital cost indicators and hospital cost variables given $\bm e_{it}$ and $c^{\text{drug}}_{it}$ is specified as:
 \begin{equation}\label{eq_hos}
\begin{split}
d^{\text{hos}}_{it} \mid \bm e_{it}, c^{\text{drug}}_{it} &\sim \text{Bernoulli}(\pi^{\text{hos}}_{ict}),\\ 
\text{logit}(\pi^{\text{hos}}_{ict}) & = \delta^{\text{hos}}_{0t} + \delta^{\text{hos}}_{1t}e^{\text{PFS}}_{it} + \delta^{\text{hos}}_{2t}e^{\text{PPS}}_{it} 
 + \delta^{\text{hos}}_{3t}\log (c^{\text{drug}}_{it}),\\[1ex]
c^{\text{hos}}_{it} \mid d^{\text{hos}}_{it}=0, \bm e_{it},c^{\text{drug}}_{it} &\sim \text{Lognormal}(\phi^{\text{hos}}_{ict}, \sigma^{\text{hos}}_{ct}),\\ 
\phi^{\text{hos}}_{ict} & = \beta^{\text{hos}}_{0t} + \beta^{\text{hos}}_{1t}e^{\text{PFS}}_{it} + \beta^{\text{hos}}_{2t}e^{\text{PPS}}_{it} 
 + \beta^{\text{hos}}_{3t}\log (c^{\text{drug}}_{it}),
\end{split}
\end{equation}
where $\pi^{\text{hos}}_{ict}$ is the probability of having $c^{\text{hos}}_{it}=0$, while $\phi^{\text{hos}}_{ict}$ and $\sigma^{\text{hos}}_{ct}$ are the mean and standard deviation parameter for $c^{\text{hos}}_{it}>0$ on the log scale. The regression parameters $\bm \delta^{\text{hos}}$ and $\bm \beta^{\text{hos}}$ capture the dependence between hospital costs, the effectiveness and the drug cost variables for the zero and non-zero components, respectively. Finally, we specify the conditional distribution of the zero adverse event cost indicators and adverse events cost variables given $\bm e_{it}$, $c^{\text{drug}}_{it}$ and $c^{\text{hos}}_{it}$ as:
 \begin{equation}\label{eq_ae}
\begin{split}
d^{\text{ae}}_{it} \mid \bm e_{it},c^{\text{drug}}_{it},c^{\text{hos}}_{it} &\sim \text{Bernoulli}(\pi^{\text{ae}}_{ict}),\\ 
\text{logit}(\pi^{\text{ae}}_{ict}) & = \delta^{\text{ae}}_{0t} + \delta^{\text{ae}}_{1t}e^{\text{PFS}}_{it} + \delta^{\text{ae}}_{2t}e^{\text{PPS}}_{it} 
 + \delta^{\text{ae}}_{3t}\log (c^{\text{drug}}_{it}) + \delta^{\text{ae}}_{4t}\log (c^{\text{hos}}_{it}),\\[1ex]
c^{\text{ae}}_{it} \mid d^{\text{ae}}_{it}=0, \bm e_{it},c^{\text{drug}}_{it},c^{\text{hos}}_{it} &\sim \text{Lognormal}(\phi^{\text{ae}}_{ict}, \sigma^{\text{ae}}_{ct}),\\ 
\phi^{\text{ae}}_{ict} & = \beta^{\text{ae}}_{0t} + \beta^{\text{ae}}_{1t}e^{\text{PFS}}_{it} + \beta^{\text{ae}}_{2t}e^{\text{PPS}}_{it} 
 + \beta^{\text{ae}}_{3t}\log (c^{\text{drug}}_{it}) + \beta^{\text{ae}}_{4t}\log (c^{\text{hos}}_{it}),
\end{split}
\end{equation}
where $\pi^{\text{ae}}_{ict}$ is the probability of having $c^{\text{ae}}_{it}=0$, while $\phi^{\text{ae}}_{ict}$ and $\sigma^{\text{ae}}_{ct}$ are the mean and standard deviation parameter for $c^{\text{ae}}_{it}>0$ on the log scale. The regression parameters $\bm \delta^{\text{ae}}$ and $\bm \beta^{\text{ae}}$ capture the dependence between adverse events costs, hospital costs, the effectiveness and the drug cost variables for the zero and non-zero components, respectively.  We note that we included the drug and hospital cost variables on the log scale in both logit and log linear regressions in Equation~\ref{eq_hos} and Equation~\ref{eq_ae} as this lead to an improvement of the model fit compared with using the costs on the original scale.
\sloppy For all parameters in the model we specify vague prior distributions: a normal distribution with a large variance on the appropriate scale for the regression parameters, e.g.~$\text{Normal}(0,10000)$, and a uniform distribution over a large positive range for the standard deviations, e.g.~$\text{Uniform}(0,10000)$. 

We note that, although the proposed model requires the specification of a relatively large number of parameters, which may make the model difficult to interpret, this, however, does not ultimately affect the final analysis, which exclusively focuses on the marginal effectiveness and cost means ($\mu_{et}$ and $\mu_{ct}$). These quantities can be retrieved by centering each variable in the effectiveness and cost modules as shown in Section~\ref{methods}. However, this procedure may become difficult to implement in practice when dealing with non-standard parametric specifications, such as the hurdle models, which make the identification of such parameters cumbersome. An alternative, although more computationally intensive, approach to overcome this problem is \textit{Monte Carlo integration}, which allows to retrieve the marginal mean estimates for each variable by randomly drawing a large number of samples from the posterior distribution of the model and then take the expectation over these sampled values as an approximation to the marginal means. 

The procedure is defined as follows. First, we fit the model to the full data, typically using some Markov Chain Monte Carlo (MCMC) methods~\textsuperscript{\citep{brooks2011handbook}} and imputing the missing values based on their predictive distributions conditional on the observed data via some data augmentation procedure~\textsuperscript{\citep{tanner1987calculation}}. Second, at each iteration of the posterior distribution, we generate a large number of samples for each component of $\bm e_{it}$ and $\bm c_{it}$ based on the posterior values for the parameters of the effectiveness and cost models in the MCMC output. Third, we approximate the posterior distribution of the marginal means by taking the expectation over the sampled values at each iteration. Finally, we derive the overall marginal means $\bm \mu= (\mu_{et},\mu_{ct})$ by summing up the marginal mean estimates for the different components of the effectiveness and costs, that is:
\begin{equation}\label{mu_ec}
\mu_{et}=\mu^{\text{PFS}}_{et} + \mu^{\text{PPS}}_{et} \;\;\; \text{and} \;\;\; \mu_{ct}=\mu^{\text{drug}}_{ct} + \mu^{\text{hos}}_{ct} + \mu^{\text{ae}}_{ct},
\end{equation}
 where $\mu^{\text{PFS}}_{et}$ and $ \mu^{\text{PPS}}_{et}$ are the pre- and post-progression mean QAS, while $\mu^{\text{drug}}_{ct}$, $\mu^{\text{hos}}_{ct}$ and $\mu^{\text{ae}}_{ct}$ are the mean for the three different components (drug, hospital and adverse events) in the TOPICAL~trial.


\subsection{Computation}\label{comp}
We fitted the model in \texttt{STAN}~\textsuperscript{\citep{carpenter2017stan}} which is a software specifically designed for the analysis of Bayesian models using a type of MCMC algorithm known as \textit{Hamiltonian Monte Carlo}~\citep{brooks2011handbook}, and which is interfaced with \texttt{R} through the package \texttt{rstan}~\textsuperscript{\citep{team2016rstan}}. Samples from the posterior distribution of the parameters of interest generated by \texttt{STAN} and saved to the \texttt{R} work space are then used to produce summary statistics and plots. We ran two chains with $15000$ iterations per chain, using a burn-in of $3000$, for a total sample of $24000$ iterations for posterior inference. For each unknown quantity in the model, we assessed convergence and autocorrelation of the MCMC simulations by using diagnostic measures including density and trace plots, the \textit{potential scale reduction factor} and the \textit{effective sample size}~\textsuperscript{\citep{Gelman2}}. A summary of the results from these convergence checks for the parameters of the model are provided in Appendix~\ref{B}, while the \texttt{STAN} code used to fit the model is provided in the supplementary material.

\subsection{Model Assessment}\label{ass}
We compute two relative measures of predictive accuracy to assess the fit of the proposed model specification (denoted as "original") with respect to a second parametric specification (denoted as "alternative"), where we replace the Gumbel distribution for $e^{\text{PFS}}_{it}$ with a Logistic distribution, the Exponential distribution for $e^{\text{PPS}}_{it}>0$ with a Weibull distribution, and the Lognormal distributions for $\bm c_{it}>0$ with Gamma distributions. We specifically rely on the \textit{Widely Applicable Information Criterion} (WAIC)~\textsuperscript{\citep{watanabe2010asymptotic}} and the \textit{Leave-One-Out Information Criterion} (LOOIC)~\textsuperscript{\citep{vehtari2017practical}}, which provide estimates for the pointwise out-of-sample prediction accuracy from a fitted Bayesian model using the log-likelihood evaluated at the posterior simulations of the parameter values. Both measures can be viewed as an improvement on the popular \textit{Deviance Information Criterion} (DIC)~\textsuperscript{\citep{spiegelhalter2002bayesian}} in that they use the entire posterior distribution, are invariant to parametrisation, and are asymptotically equal to Bayesian cross-validation~\textsuperscript{\citep{gelman2014understanding}}. These information criteria are obtained based on the model deviance and a penalty for model
complexity known as effective number of parameters ($p_{D}$ ) and, when comparing a set of models based on the same data, the one associated with the lowest WAIC or LOOIC is the best-performing, among those assessed. 

Results between the two alternative specifications are reported in Table~\ref{tab_ic}. 
\begin{table}[!h]
\centering
TABLE 1
\end{table}
For both criteria, the values associated with the "original" specification of the model are systematically lower compared with those from the "alternative" parameterisation, and result in an overall better fit to the data for the first model. We have explored different configurations of distribution assignment among those shown in Table~\ref{tab_ic} for the effectiveness and cost variables (including also the Normal distribution which, however, was always associated with the worst fit). The results from these comparisons suggest that the "original" specification of the model remains the one associated with the best performance.

We additionally assess the absolute fit of the model using the observed and replicated data. The latter are generated from the posterior predictive distributions of the models using the posterior samples of the parameters indexing the distributions of each effectiveness and cost variable. We use the posterior estimates from these parameters to jointly sample $10000$ replications of the data which are then used for model assessment. We computed different types of graphical posterior predictive checks, either in terms of the entire distributions using density and cumulative density plots, or in terms of the marginal mean estimates between the real and replicated data, which are provided in Appendix~\ref{D}. Overall these checks suggest a relatively good fit of the model for each modelled variable.

\section{Results}\label{results}
This section discusses the results of the model from a twofold perspective: first, focus is given to summarising the posterior distribution of the main quantities of interest, namely the marginal means of each component of the effectiveness ($\mu^{\text{PFS}}_{et},\mu^{\text{PPS}}_{et}$) and cost ($\mu^{\text{drug}}_{ct},\mu^{\text{hos}}_{ct},\mu^{\text{ae}}_{ct}$) variables and the marginal means for the aggregated variables, namely $\bm \mu=(\mu_{et},\mu_{ct})$; second, the economic results are discussed by computing the probability that the new intervention is cost-effective with respect to the control. 

\subsection{Posterior Estimates}\label{post_est}
Figure~\ref{fig_means} compares the posterior means (squares) together with the $50\%$ (thick lines) and $95\%$ (thin lines) highest posterior density (HPD) credible intervals for the marginal means of each effectiveness and cost components, obtained from fitting the model to all cases under a MAR assumption. Results associated with the control ($t=1$) and intervention ($t=2$) group are indicated with red and blue colours, respectively.
\begin{figure}[!h]
\centering
FIGURE 4
\end{figure}
Posterior QAS means are on average higher for the PFS compared with the PPS component in both treatment groups. However, both $50\%$ and $95\%$ HPD intervals suggest that the estimates associated with the intervention group have a much higher degree of variability compared with those from the control, especially for the PPS component. The posterior cost means for each component show that the intervention group is associated with systematically higher values with respect to the control, especially for the drug costs which by far cover most of the total costs in the intervention. HPD intervals for mean costs show a relatively high degree of skewness with posterior mean estimates being closer to the upper bounds of the $50\%$ intervals compared to the lower bounds.

We derived the aggregated QAS and cost means for each treatment group $(\mu_{et}, \mu_{ct})$ by summing up the posterior mean estimates of the different components for each type of variable. We then computed the incremental mean estimates between the two groups, denoted with $\Delta_e=\mu_{e2}-\mu_{e1}$ and $\Delta_c=\mu_{c2}-\mu_{c1}$, together with the Incremental Cost Effectiveness Ratio (ICER), representing the cost per QAS gained between the two interventions. Table~\ref{tab_agg_means} shows selected posterior summaries, including means, medians, standard deviations and $95\%$ HPD intervals, for the marginal and incremental mean estimates associated with the two intervention groups.
\begin{table}[!h]
\centering
TABLE 2
\end{table}
Overall, the posterior results indicate that the new intervention is associated with systematically higher QAS and costs compared to the control, with a positive mean QAS increment of $0.17$, a positive mean cost increment of $\pounds 12602$, and with $95\%$ intervals that exclude zero for both quantities. We note that posterior estimates for the marginal means in the control group show a considerably lower degree of variability (standard deviations of $0.02$ and $\pounds 424$) compared with those from the intervention group (standard deviations of $0.05$ and $\pounds 2628$). Finally, the additional cost per unit of QAS gained is estimated to be roughly $\pounds 80000$ for $t=2$ compared to $t=1$.

\subsection{Economic Evaluation}\label{econ_eval}
We complete the analysis by assessing the probability of cost-effectiveness for the new intervention with respect to the control. A general advantage of using a Bayesian approach is that the economic analysis can be easily performed without the need to use ad-hoc methods to represent uncertainty around point estimates (e.g.~bootstrapping). Indeed, once the statistical model is fitted to the data, the samples from the posterior distributions of the parameters of interest can be used to compute different types of summary measures of cost-effectiveness. 

We specifically rely on the examination of the cost-effectiveness plane (CEP)~\textsuperscript{\citep{Black}} and the cost-effectiveness acceptability curve (CEAC)~\textsuperscript{\citep{VanHout}} to summarise the economic analysis. Figure~\ref{ee_plot} (panel a) shows the plot of the CEP, which is a graphical representation of the joint distribution for the mean effectiveness and costs increments between the two treatment groups. The slope of the straight line crossing the plane is the "willingness-to-pay’" threshold (which is often indicated as $k$). This can be considered as the amount of budget that the decision maker is willing to spend to increase the health outcome of $1$ unit and effectively is used to trade clinical benefits for money. Current recommendations for generic interventions suggest a value of $k$ between $\pounds 20000 - 30000$; however, for end of life treatments, such as cancer treatments, the recommended threshold values are typically higher and lie in a range between $\pounds 50000 - 60000$ or above~\textsuperscript{\citep{NICE2013}}. Points lying below this straight line fall in the so-called \textit{sustainability area}~\textsuperscript{\citep{Baioa}} and suggest that the new intervention is more cost-effective than the control. 
\begin{figure}[!h]
\centering
FIGIRE 5
\end{figure}
Almost all samples fall in the north-east quadrant of the plane. This suggests that the intervention is likely to be more effective and more expensive compared with the control. At $k=\pounds 55000$, the ICER (and the majority of the samples) falls outside the sustainability area, and therefore indicates that the intervention is unlikely to be considered cost-effective at the chosen value of $k$. Figure~\ref{ee_plot} (panel b) shows the CEAC, which is obtained by computing the proportion of points lying in the sustainability area on varying the willingness-to-pay threshold $k$. The CEAC estimates the probability of cost-effectiveness, thus providing a simple summary of the uncertainty that is associated with the "optimal" decision making that is suggested by the ICER. The graph shows how, starting from null to relatively small probabilities of cost-effectiveness for $k<\pounds 50000$, as the value of the willingness to pay threshold is increased, the chance that the new intervention becomes cost-effective rises up to near full certainty for $k>\pounds 150000$. We note that these cost-effectiveness conclusions should be interpreted with care, due to limitations associated with the data analysed which are explicitly discussed in Section~\ref{discussion} (e.g.~based on a subset of the original data). However, this does not affect the generalisability of our framework, which implements a flexible modelling approach for handling the typical characteristics of trial-based partitioned survival economic data.

\section{Discussion}\label{discussion}
In this paper, we have proposed a general framework for partitioned survival cost-utility analysis using patient level data (e.g.~from a trial), which takes into account the correlation between costs and effectiveness, skewness in the distribution of the observed data, the presence of structural zeros, and missing data. Although alternative approaches have been proposed in the literature to handle the statistical issues affecting cost-effectiveness data, they had either considered only some of these issues separately~\textsuperscript{\citep{OHaganb,Basu,baio2014bayesian}} or did not specifically focus on partitioned survival analyses~\textsuperscript{\citep{gabrio2019full, gabrio2019bayesian}}. The approach developed in Section~\ref{methods} uses a flexible structure and allows the distributions of different components of the effectiveness and costs to be modelled so to handle the typical idiosyncrasies that affect each variable, all within a joint probabilistic framework. This is a key advantage of the Bayesian approach compared with other approaches, especially in health economic evaluations where the main objective is not statistical inference per se, but rather to assess the uncertainty in the decision-making process, induced by the uncertainty in the model inputs~\textsuperscript{\citep{claxton1999irrelevance,Claxton2005}}. 

The economic results from our case study should be interpreted with caution and some potential limitations in terms of the generalisability of the proposed framework to other studies should be highlighted. First, our analysis of TOPICAL is based on a subset of the individuals in the original trial (made available to us) and therefore it is difficult to draw any cost-effectiveness conclusions about the trial from this analysis. Second, although the results are obtained under a MAR assumption, which is typically considered more plausible than just focussing on the complete cases, missingness assumptions can never be checked from the data at hand. It is possible that the assumption of MAR is not tenable, which may therefore introduce some bias. It is generally recommended that departures from MAR are explored in sensitivity analysis to assess the robustness of the conclusions to some plausible MNAR scenarios~\textsuperscript{\citep{Daniels}}. However, given the limitations of our analysis in terms of the interpretation of the trial results and the lack of any external information to guide the choice of the MNAR departures, we decided not to pursue such analyses here. We note that different approaches are available to conduct sensitivity analysis to MNAR, some of which can be implemented within a Bayesian framework, for example through the elicitation of expert's opinions using prior distributions~\textsuperscript{\citep{Daniels,mason2018bayesian}}. Finally, while the computation of QAS based on utility and survival data from TOPICAL can be assumed to be valid (since $>99\%$ of individuals in the trial have progressed/died by the time of the analysis), it is generally known that this calculation may be invalid if the survival time for some patients is not observed, as it may introduce informative censoring which distorts the inferences~\textsuperscript{\citep{glasziou1990quality}}. Alternative approaches have been proposed to overcome this problem, which mostly rely on the separate calculation of population-level summary measures for the utility and survival components (e.g.~using linear mixed models for the utilities and Kaplan-Meier estimates for the survival times) which are then combined together to generate QAS estimates~\textsuperscript{\citep{khan2015design}}.

In conclusions, although our approach may not be applicable to all cases, the data analysed are very much representative of the “typical” data used in partitioned survival cost-utility analysis alongside clinical trials. Thus, it is highly likely that the same features apply to many real cases. This is a very important, if somewhat overlooked problem, as methods that do not take into account the complexities affecting patient-level data, while being easier to implement and well established among practitioners, may ultimately mislead cost-effectiveness conclusions and bias the decision-making process.

\ack{We wish to thank the UCL CRUK Cancer Trials Centre for providing a subset of data from the TOPICAL trial.}

\bibliographystyle{unsrtnat}
\bibliography{model_survivalCUA_ref}

\clearpage

\begin{appendices}

\section{Missingness Patterns in TOPICAL}\label{A1}

\begin{table}[!h]
\centering
TABLE 3
\end{table}

\section{MCMC Convergence}\label{B}
This section reports some summary diagnostics of the MCMC algorithm, including density and trace plots, for all key parameters of the model fitted to the TOPICAL data. For all parameters, no substantial issue in convergence is detected. All plots were generated from \texttt{R} using functions in the package \texttt{bayesplot}~\textsuperscript{\citep{gabry2018bayesplot}}, which provides a variety of tools to perform diagnostic and posterior predictive checks for Bayesian models. 

\begin{figure}[!h]
\centering
FIGURE 6
\end{figure}

\begin{figure}[!h]
\centering
FIGURE 7
\end{figure}

\begin{figure}[!h]
\centering
FIGURE 8
\end{figure}

\begin{figure}[!h]
\centering
FIGURE 9
\end{figure}

\section{Posterior Predictive Checks}\label{D}
This section reports some graphical posterior predictive checks, including empirical density, cumulative density and mean plots, for all variables of the model fitted to each treatment group in the TOPICAL trial. In all cases, no substantial discrepancy between the observed and predicted data is detected which suggests a generally good fit of the model. All plots were generated from \texttt{R} using functions in the package \texttt{bayesplot}~\textsuperscript{\citep{gabry2018bayesplot}}, which provides a variety of tools to perform diagnostic and posterior predictive checks for Bayesian models. 

\begin{figure}[!h]
\centering
FIGURE 10
\end{figure}

\begin{figure}[!h]
\centering
FIGURE 11
\end{figure}

\begin{figure}[!h]
\centering
FIGURE 12
\end{figure}

\begin{figure}[!h]
\centering
FIGIRE 13
\end{figure}

\begin{figure}[!h]
\centering
FIGURE 14
\end{figure}

\end{appendices}

\clearpage

\begin{figure}[!h]
\centering
\includegraphics[scale=1]{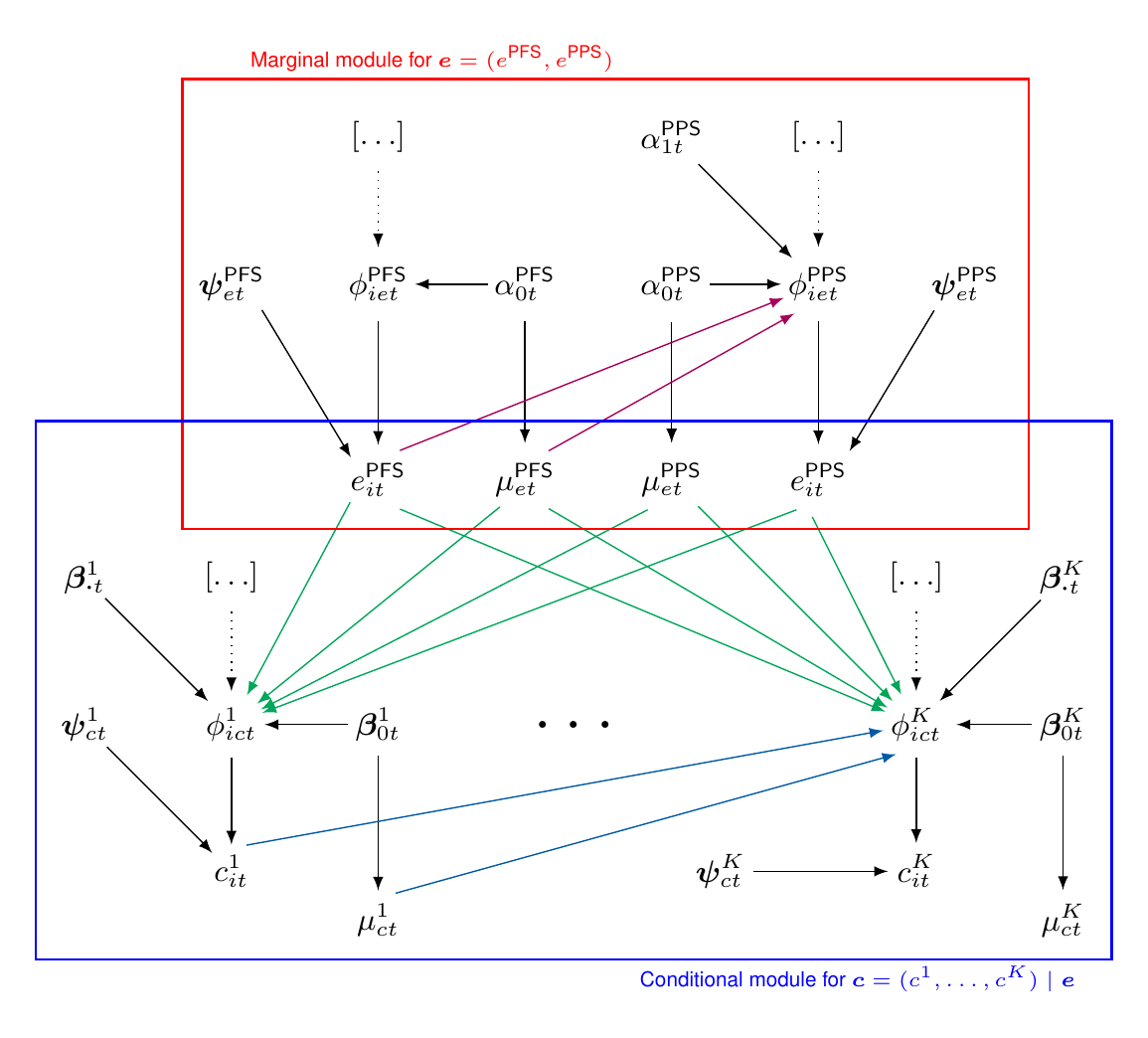}
\caption{Joint distribution $p(\bm e, \bm c)$, expressed in terms of a marginal distribution for the effectiveness variables $\bm e=(e^{\text{PFS}},e^{\text{PPS}})$ and a conditional distribution for the cost variables $\bm c =(c^1,\ldots,c^K)$ given $\bm e$, respectively indicated with a solid red and blue box. The parameters indexing the corresponding distributions or modules are denoted with different Greek letters, whereas $i$ and $t$ denote the individual and treatment indices. The notation $\bm \beta^{1}_{\cdot t}$ and $\bm \beta^{K}_{\cdot t}$ indicates the set of the conditional mean cost regression parameters for $c^1$ and $c^K$, excluding the intercepts. The solid black and coloured arrows show the dependence relationships between the parameters within and between different modules, respectively. The three large dots indicate the inclusion in the framework of the conditional distributions for the cost variables $c^k \mid \bm e, c^{k},\ldots,c^{k-1}$, for $2<k<K$, omitted for clarity from the figure, while the small dots enclosed in the square brackets indicate the potential inclusion of other covariates at the mean level in each module.}\label{model_frame}
\end{figure}

\begin{figure}[!h]
\centering
\subfloat[control]{\includegraphics[scale=0.3]{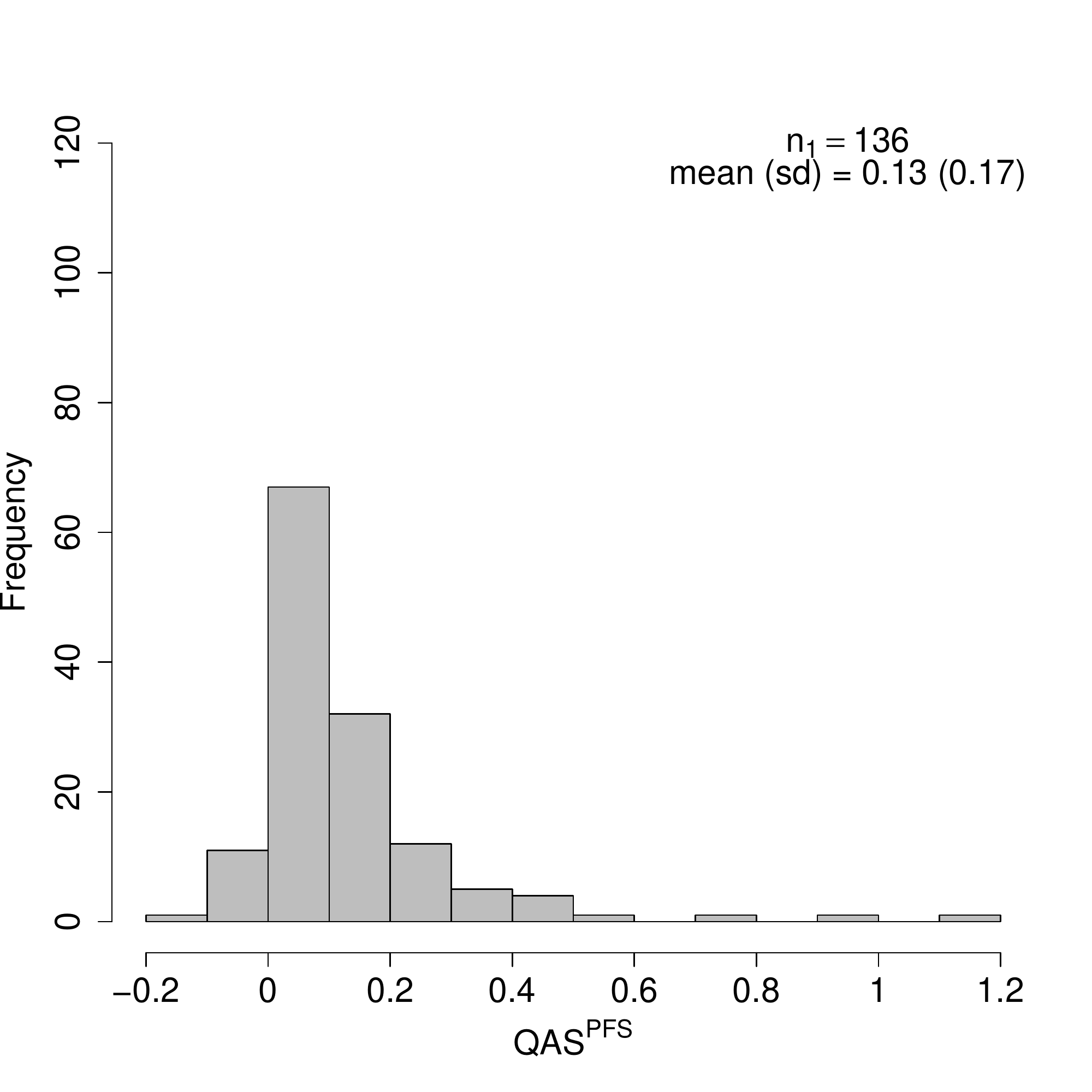}}
\subfloat[control]{\includegraphics[scale=0.3]{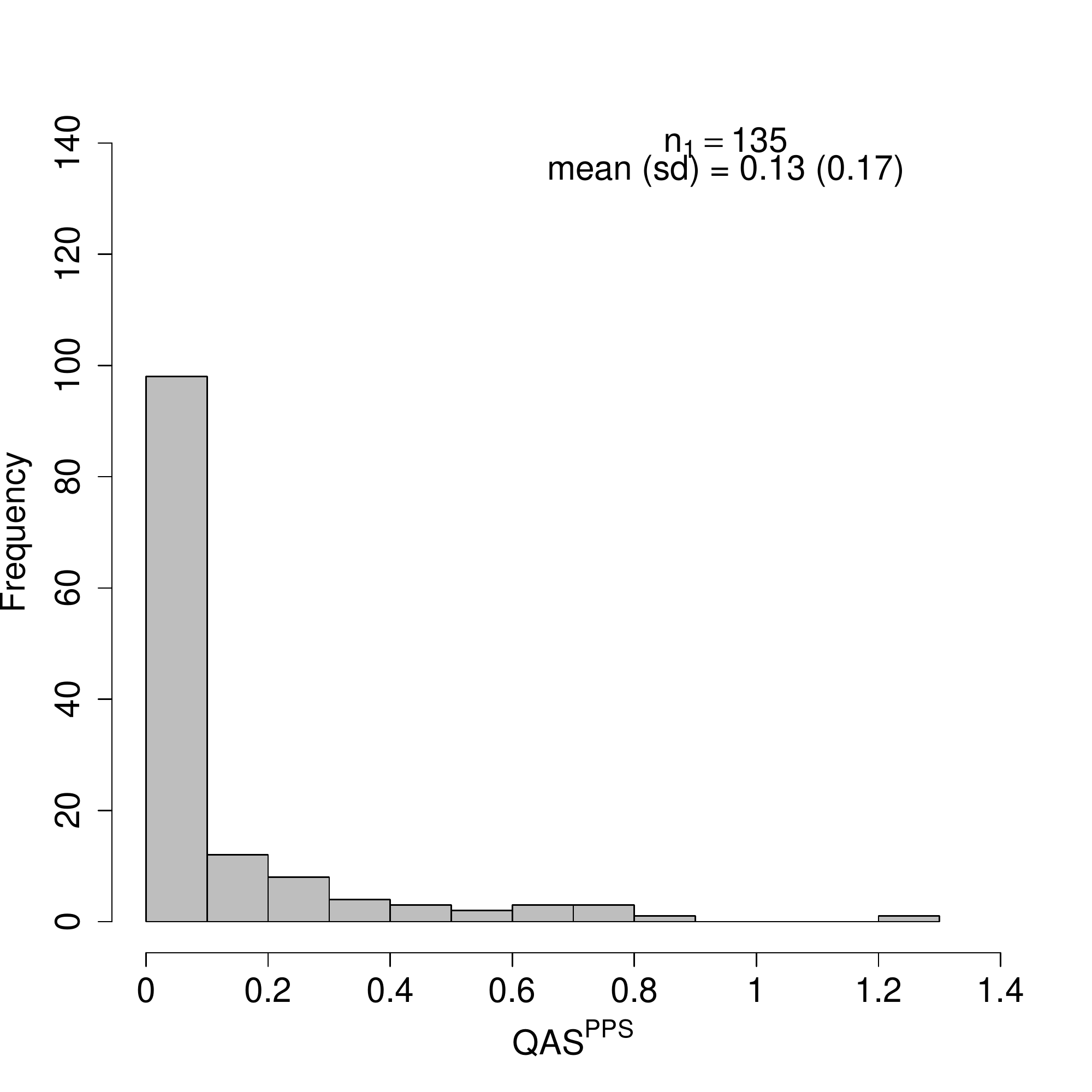}}\\
\subfloat[intervention]{\includegraphics[scale=0.3]{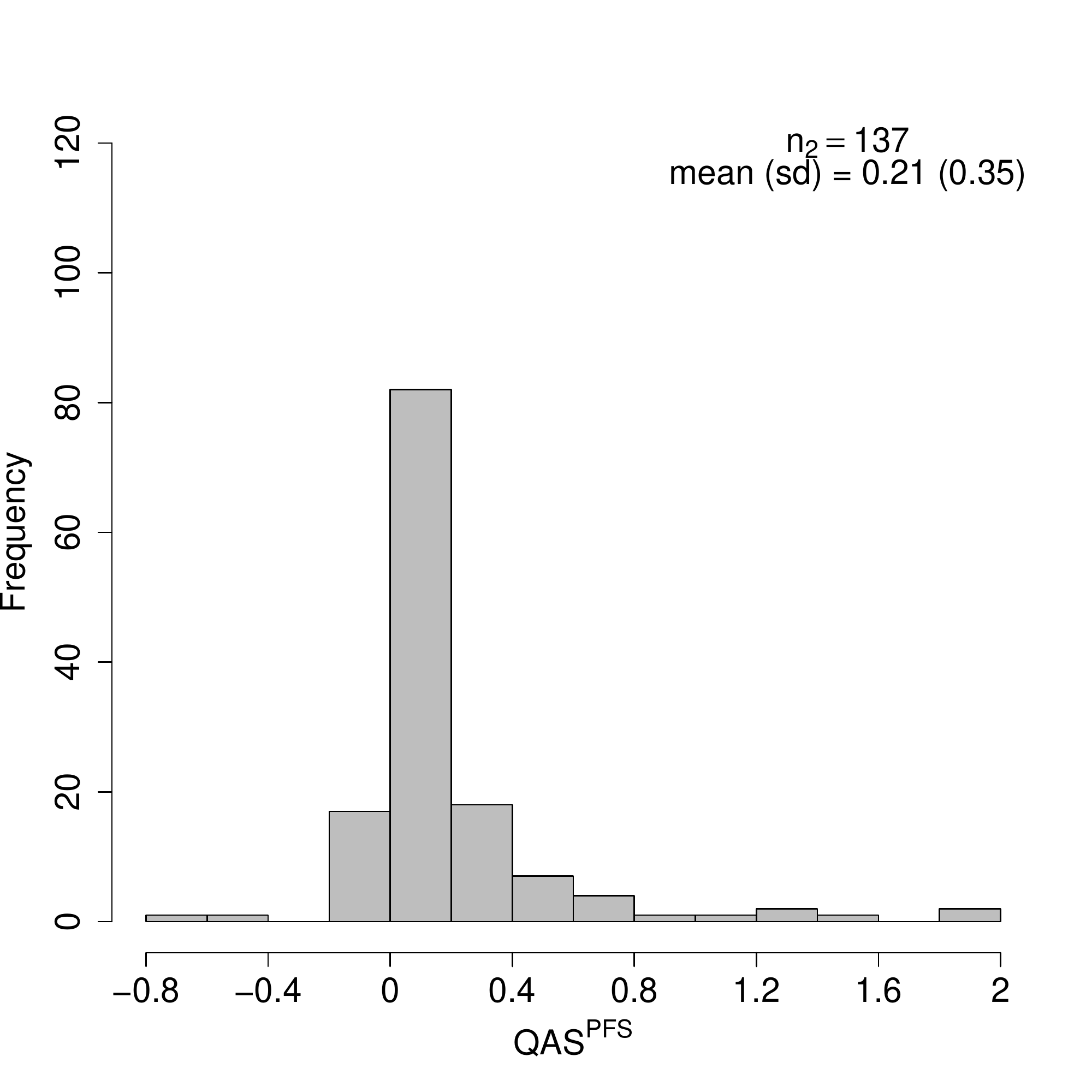}}
\subfloat[intervention]{\includegraphics[scale=0.3]{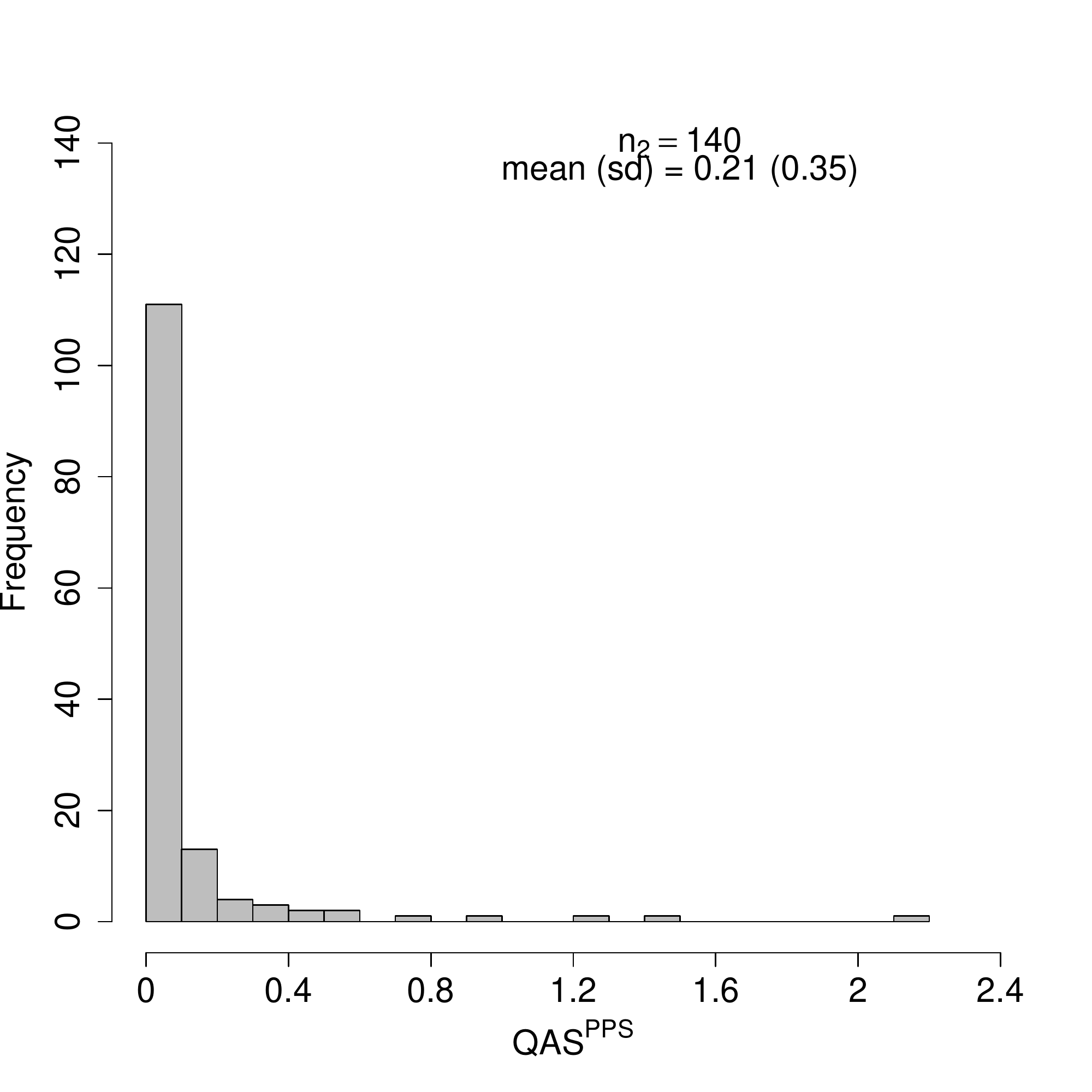}}
\caption{Histograms of the distributions of the pre- and post-progression QAS data, in the control (a-b) and intervention (c-d) group. For both variables and in both arms, skewness of the observed data is~apparent.}\label{hist_e}
\end{figure}

\begin{figure}[!h]
\centering
\subfloat[control]{\includegraphics[scale=0.3]{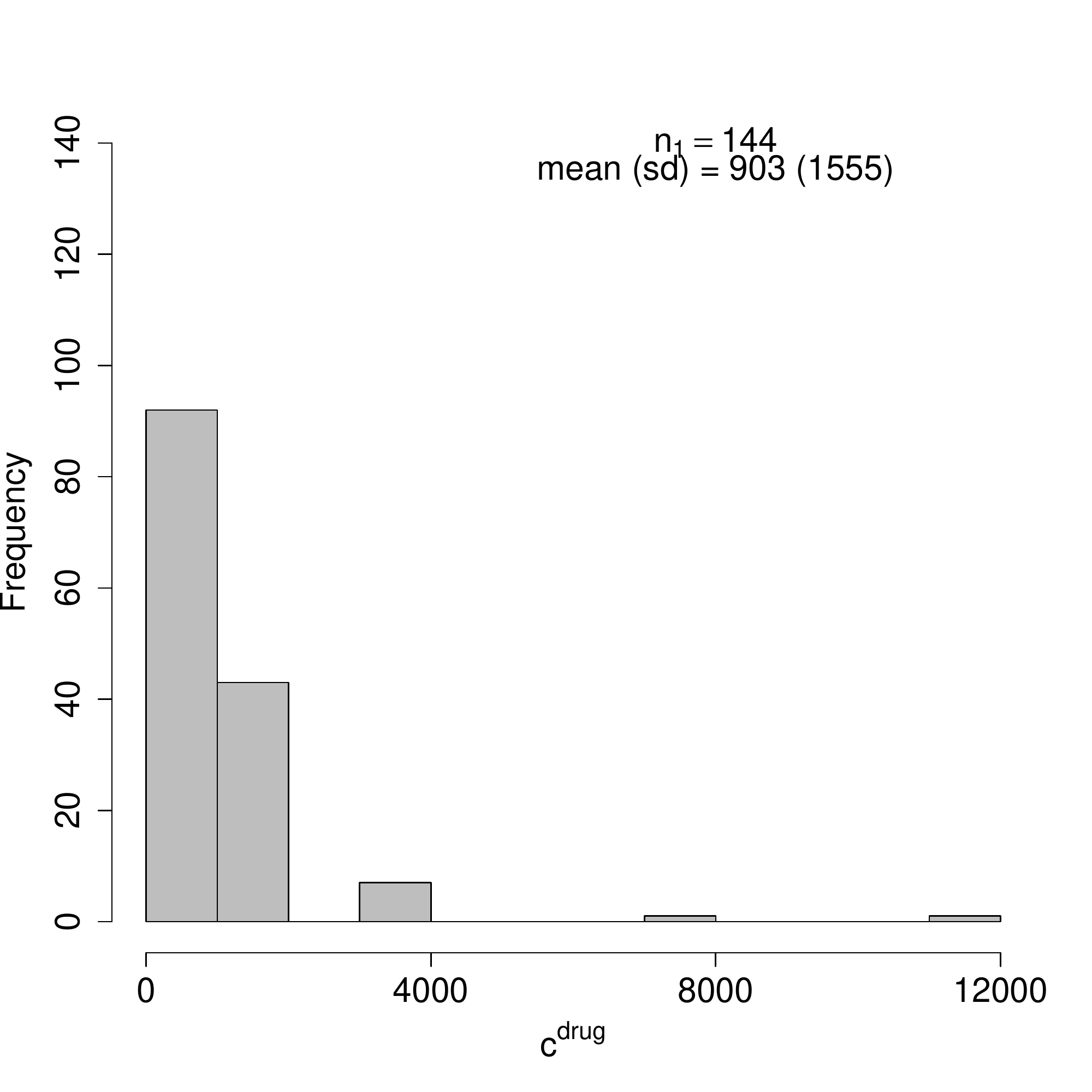}}
\subfloat[control]{\includegraphics[scale=0.3]{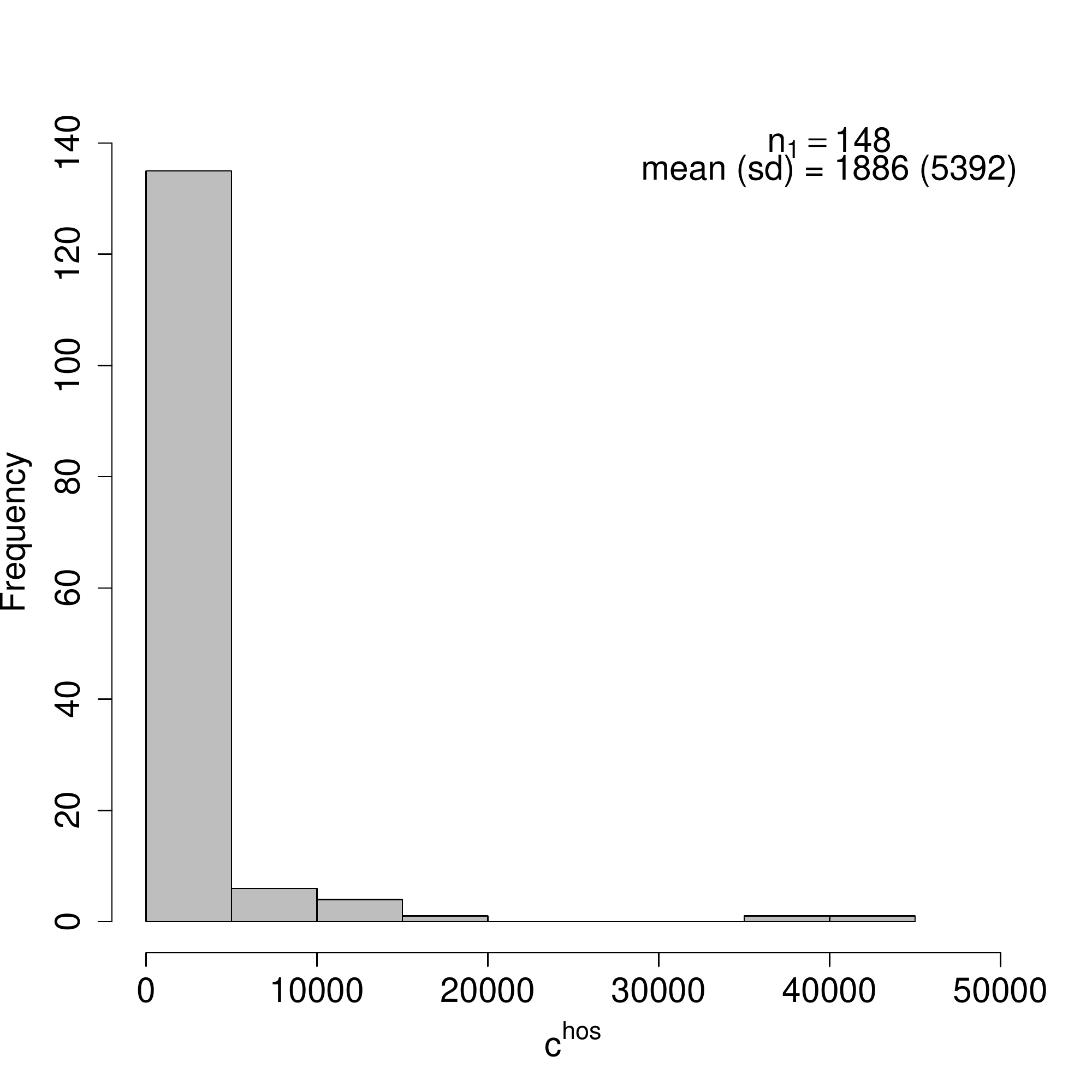}}
\subfloat[control]{\includegraphics[scale=0.3]{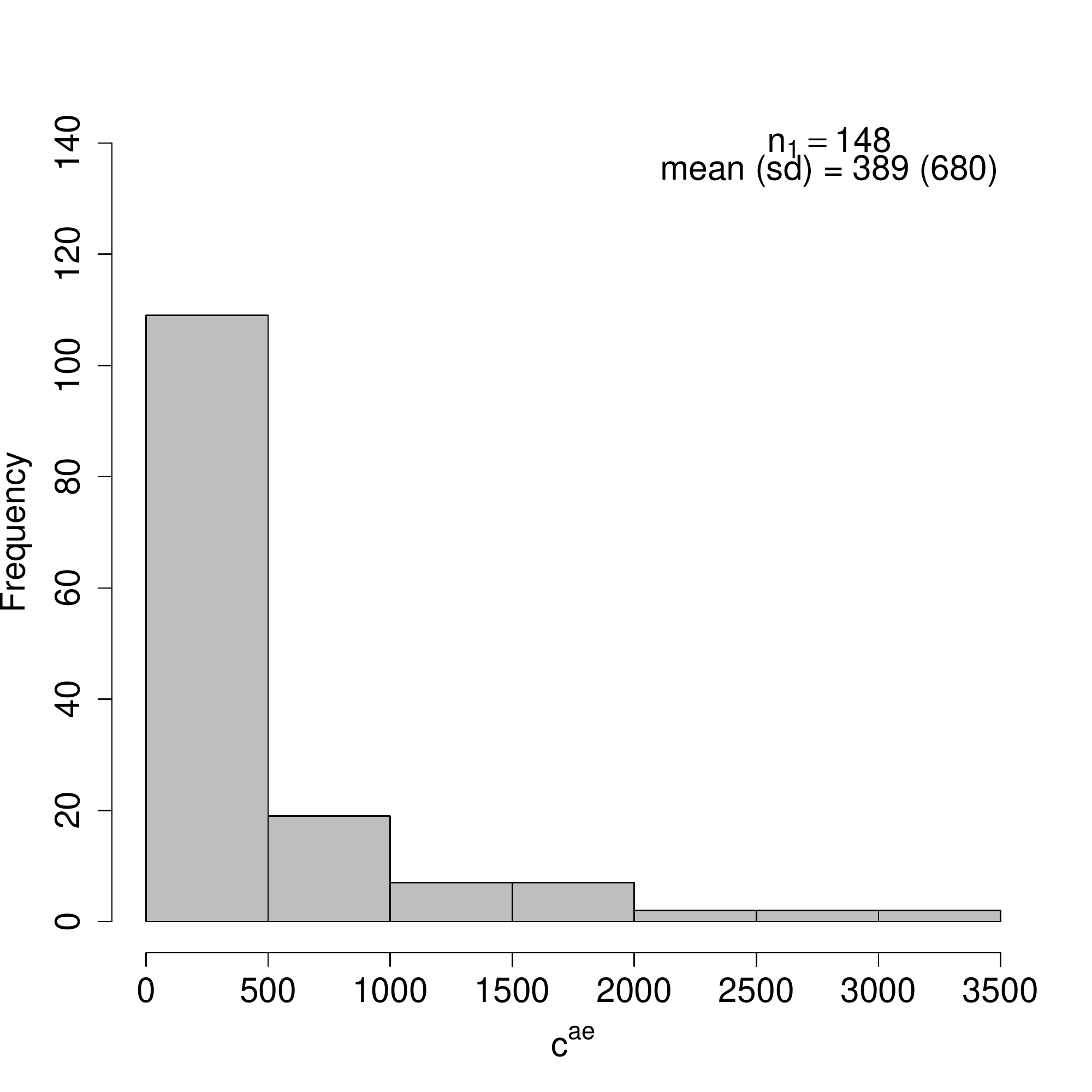}}\\
\subfloat[intervention]{\includegraphics[scale=0.3]{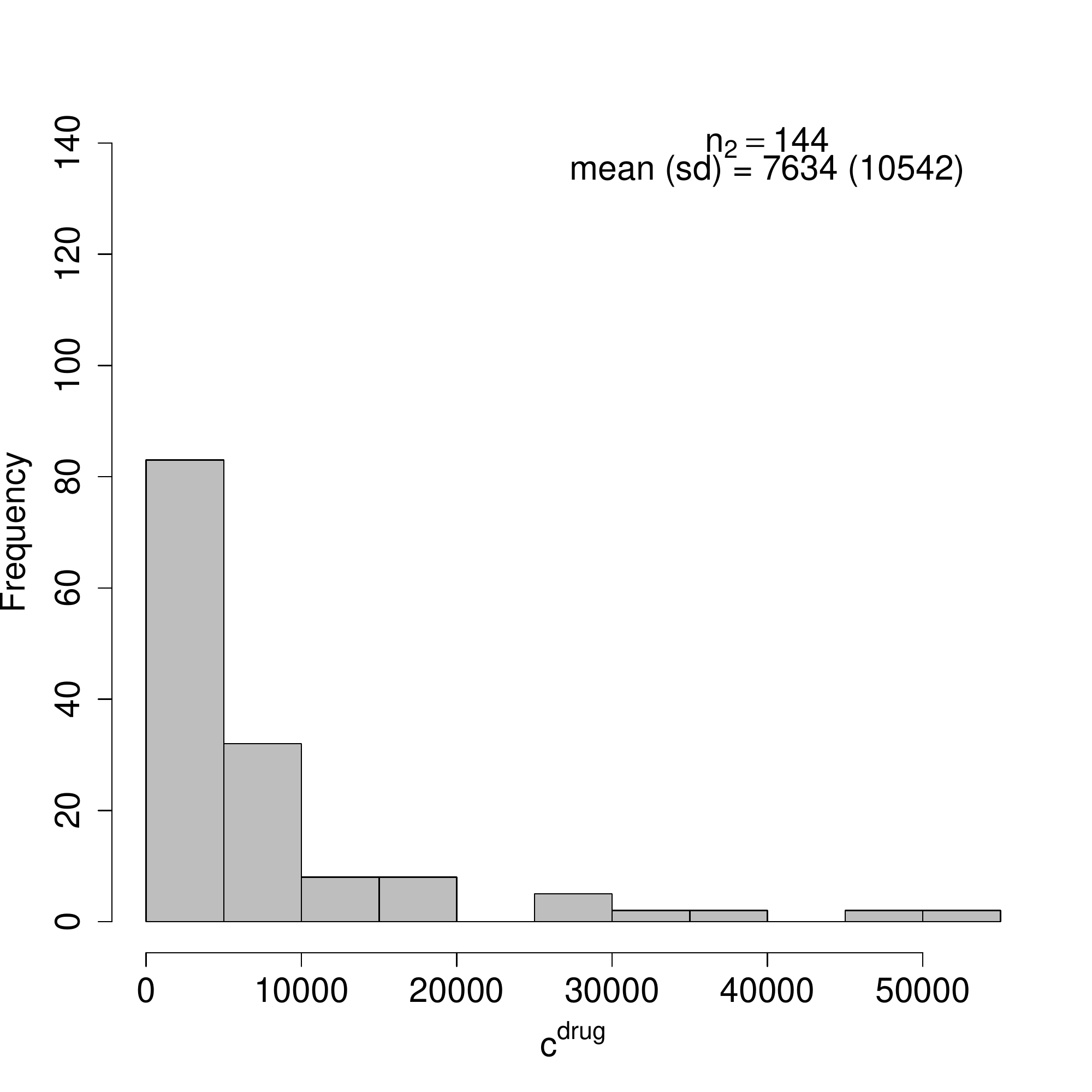}}
\subfloat[intervention]{\includegraphics[scale=0.3]{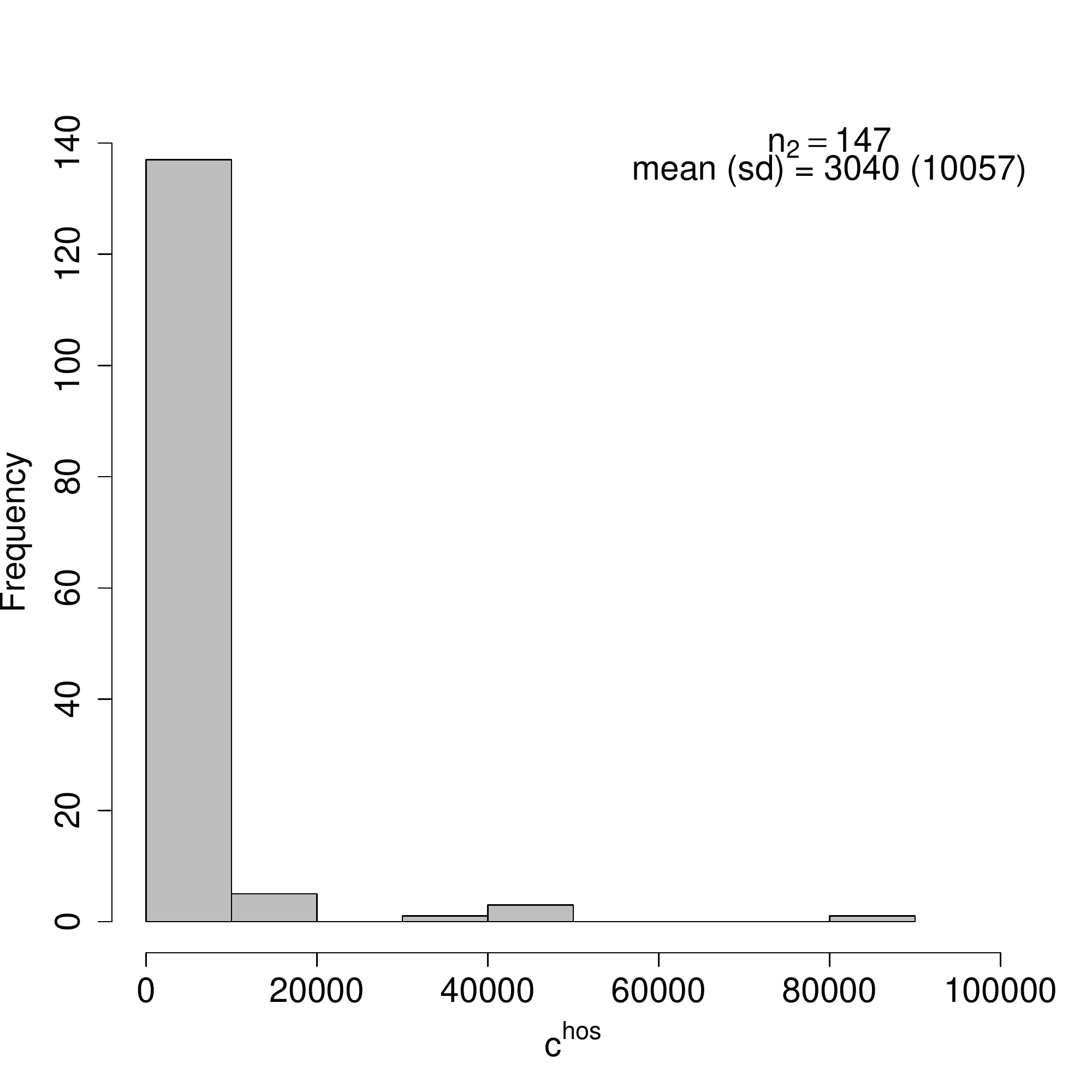}}
\subfloat[intervention]{\includegraphics[scale=0.3]{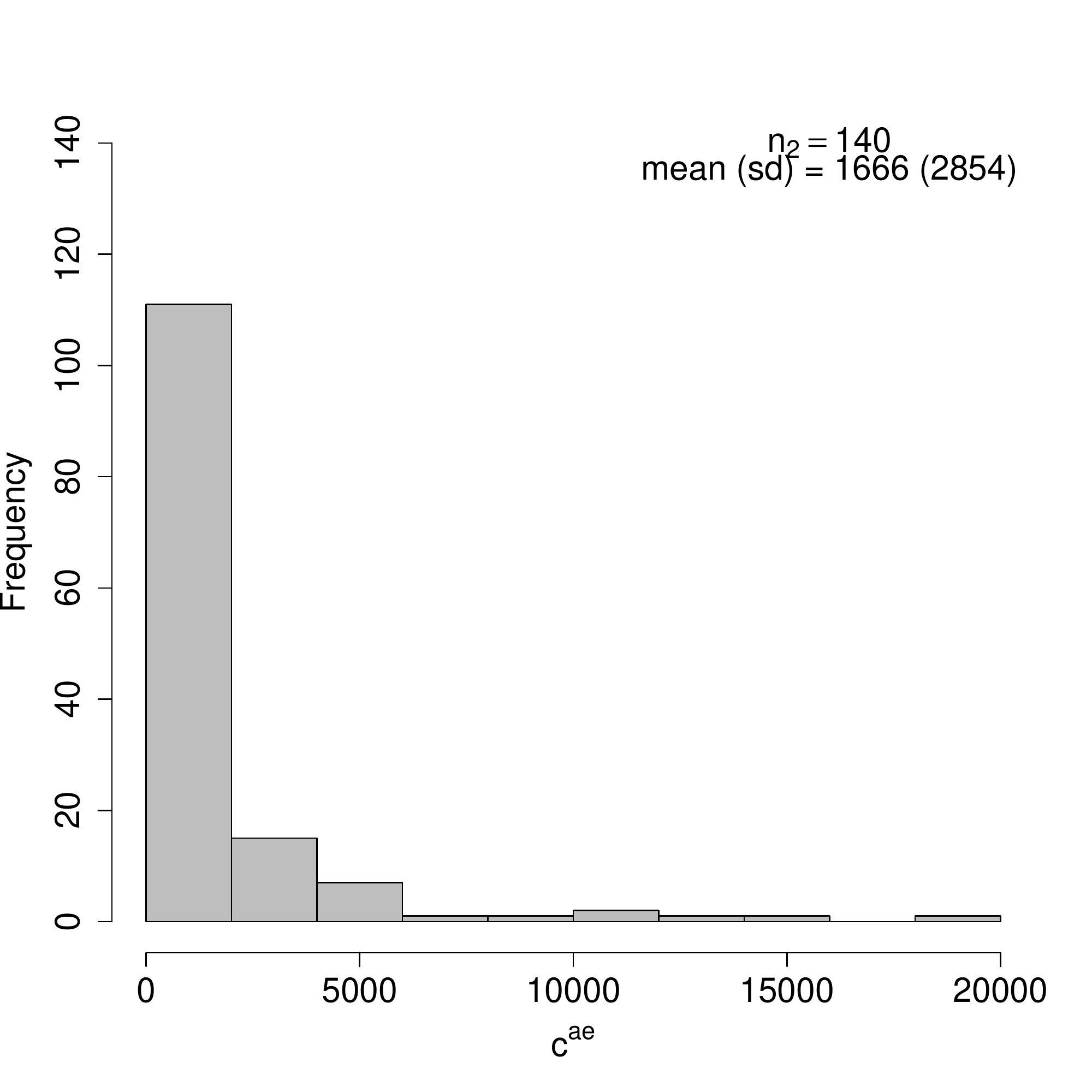}}
\caption{Histograms of the distributions of the three cost components (drug, hospital and adverse events) in the control (a-c) and intervention (d-f) groups (all costs are expressed in pounds). For all variables and in both arms, skewness of the observed data is~apparent.}\label{hist_c}
\end{figure}

\begin{table}[!h]
\centering
\scalebox{0.9}{
\begin{tabular}{c|ccc|ccc}
\toprule
& \multicolumn{3}{c|}{Original} &\multicolumn{3}{c}{Alternative}\\
\midrule
variable & Distribution & WAIC($p_D$) & LOOIC($p_D$) & Distribution &  WAIC($p_D$) & LOOIC($p_D$)  \\ 
\midrule
$e^{\text{PFS}}$ & Gumbel &  -109(11) & -107(12)  & Logistic & -68(8) & -68(8) \\ 
    $e^{\text{PPS}} \mid e^{\text{PFS}}$ & Exponential & 34(10)  &  35(10)  & Weibull & 36(8) & 38(9) \\ 
    $c^{\text{drug}} \mid \bm e$ & Lognormal & 3283(16)  & 3286(17)  &  Gamma &  3361(26) &  3365(28) \\ 
   $c^{\text{hos}} \mid \bm e, c^{\text{drug}}$ & Lognormal &  3437(15)  & 3438(15)  &  Gamma & 3659(15)  & 3660(16)  \\ 
    $c^{\text{ae}} \mid \bm e,c^{\text{drug}},c^{\text{hos}}$& Lognormal &  3208(22) & 3211(23)  &  Gamma & 3437(38)  & 3433(36)  \\ 
  \midrule
 Total & & 9853(74)  & 9863(77)  & &  10425(95) & 10428(97)\\
\bottomrule
\end{tabular}
}
\caption{\label{tab_ic} WAIC, LOOIC and effective number of parameter ($p_D$ ) estimates for each variable in the model. The "original" and "alternative" model specifications are assessed using different distributions for the pre-/post-progression QAS and the cost data. Total WAIC, LOOIC and $p_D$ values are reported at the bottom of the table.}
\end{table}

\begin{figure}[!h]
\centering
\subfloat[]{\includegraphics[scale=0.4]{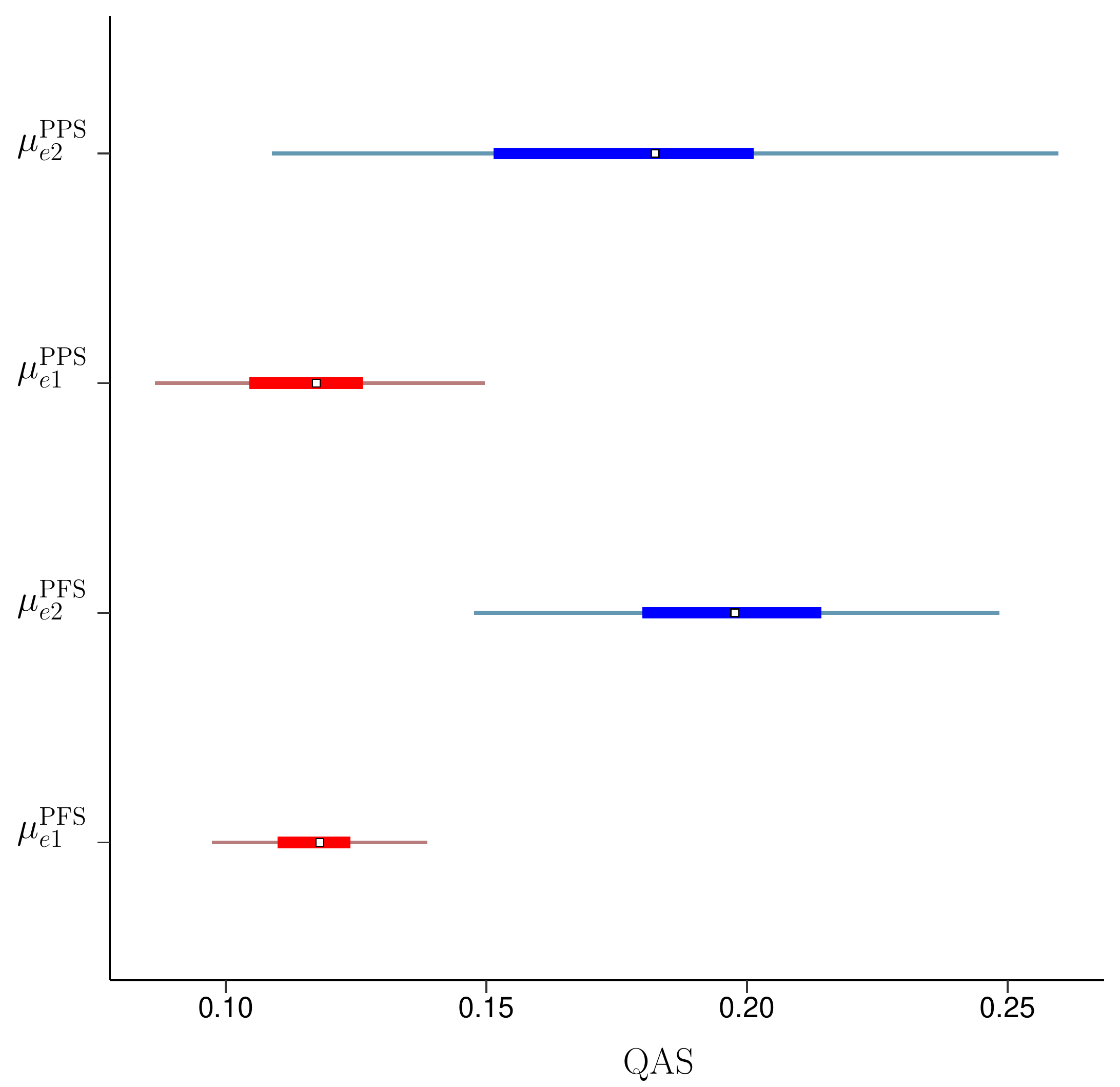}}
\subfloat[]{\includegraphics[scale=0.4]{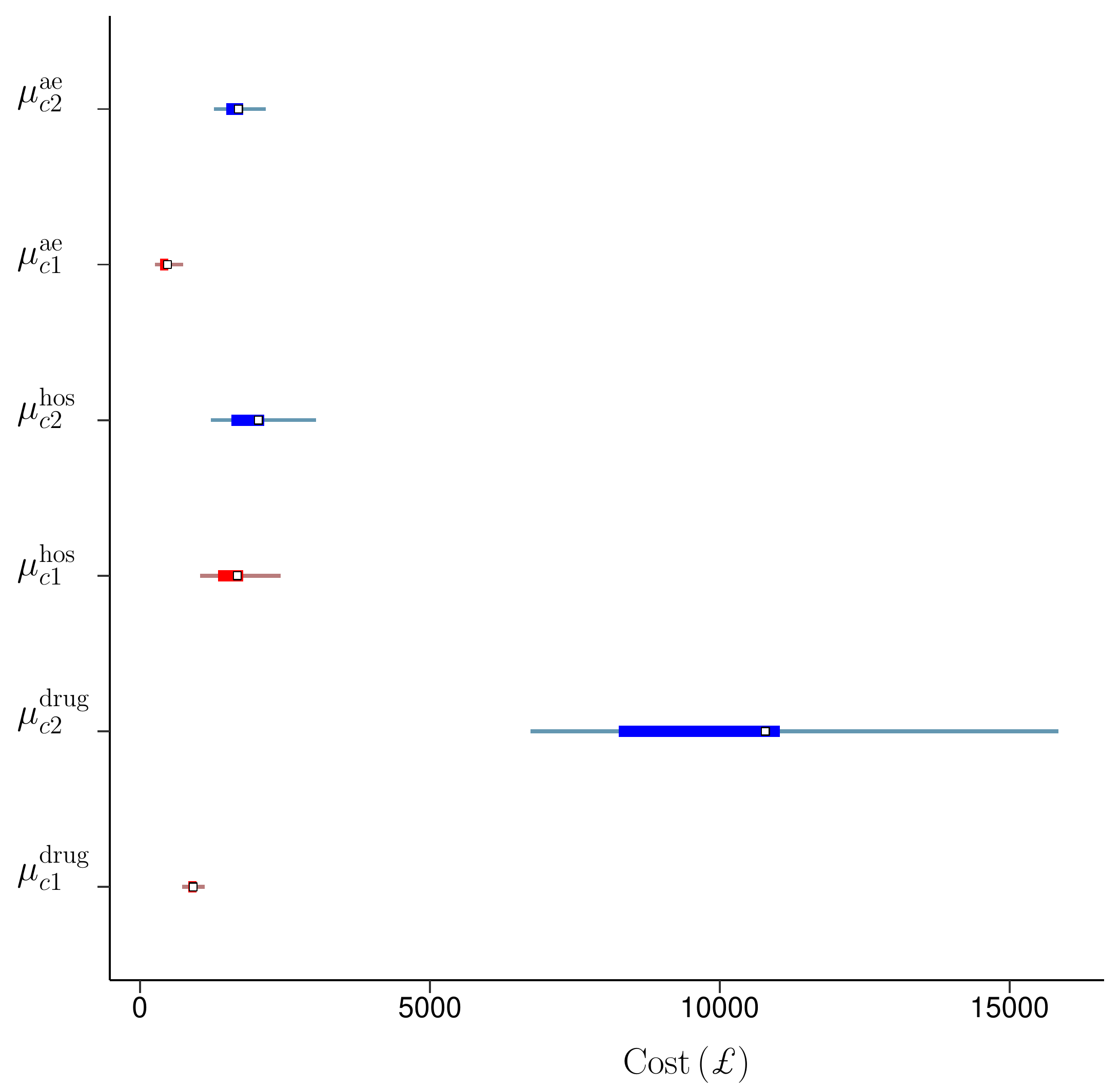}}
\caption{Posterior means (squares), $50\%$ (thick lines) and $95\%$ (thin lines) HPD credible intervals for the marginal means of pre- and post-progression QAS (panel a) and for the marginal means of the drug, hospital and adverse events cost (panel b) in the control (red) and in the intervention (blue) group in the TOPICAL~trial.}\label{fig_means}
\end{figure}

\begin{table}[!h]
\centering
\begin{tabular}{cccccc}
  \toprule
 \textbf{Parameter} & \textbf{mean} & \textbf{median} & \textbf{sd} & \multicolumn{2}{c}{\textbf{95\% CI}} \\ 
  \midrule
\textbf{Control $(t=1)$} &  &  &  &  &  \\ 
$\mu_{e1}$ & 0.24 & 0.23 & 0.02 & 0.20 & 0.27 \\ 
  $\mu_{c1}$ & 3059 & 3001 & 424 & 2329 & 3898 \\ [2ex]
\textbf{Intervention $(t=2)$} &  &  &  &  &  \\ 
 $\mu_{e2}$ & 0.38 & 0.38 & 0.05 & 0.29 & 0.47 \\ 
  $\mu_{c2}$ & 14519 & 14055 & 2628 & 10235 & 19681 \\ [2ex]
\textbf{Incremental} &  &  &  &  &  \\ 
  $\Delta_e$ & 0.14 & 0.14 & 0.05 & 0.05 & 0.24 \\ 
  $\Delta_c$ & 11460 & 11013 & 2666 & 7282 & 16983 \\ [1ex]
  ICER & 79233 &  &  &  &  \\ 
  \bottomrule
\end{tabular}
\caption{Posterior means, medians, standard deviations and $95\%$ HPD credible intervals for the marginal ($\mu_{et},\mu_{ct}$) and incremental ($\Delta_e,\Delta_c$) mean total QAS and cost estimates associated with the control ($t=1$) and intervention ($t=2$) group in the TOPICAL trial. Costs are expressed in $\pounds$.}\label{tab_agg_means}
\end{table}

\begin{figure}[!h]
\centering
\subfloat[]{\includegraphics[scale=0.4]{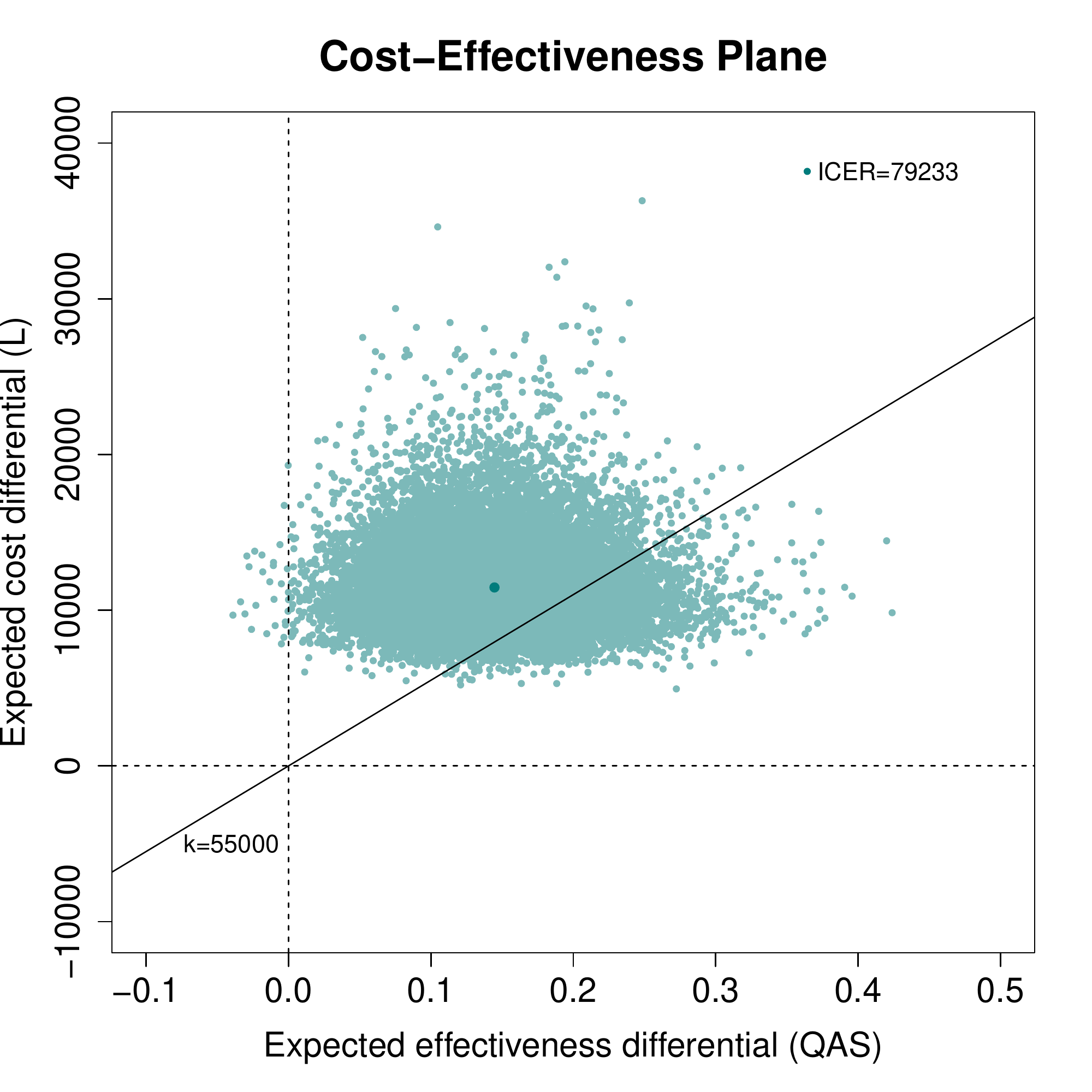}}
\subfloat[]{\includegraphics[scale=0.4]{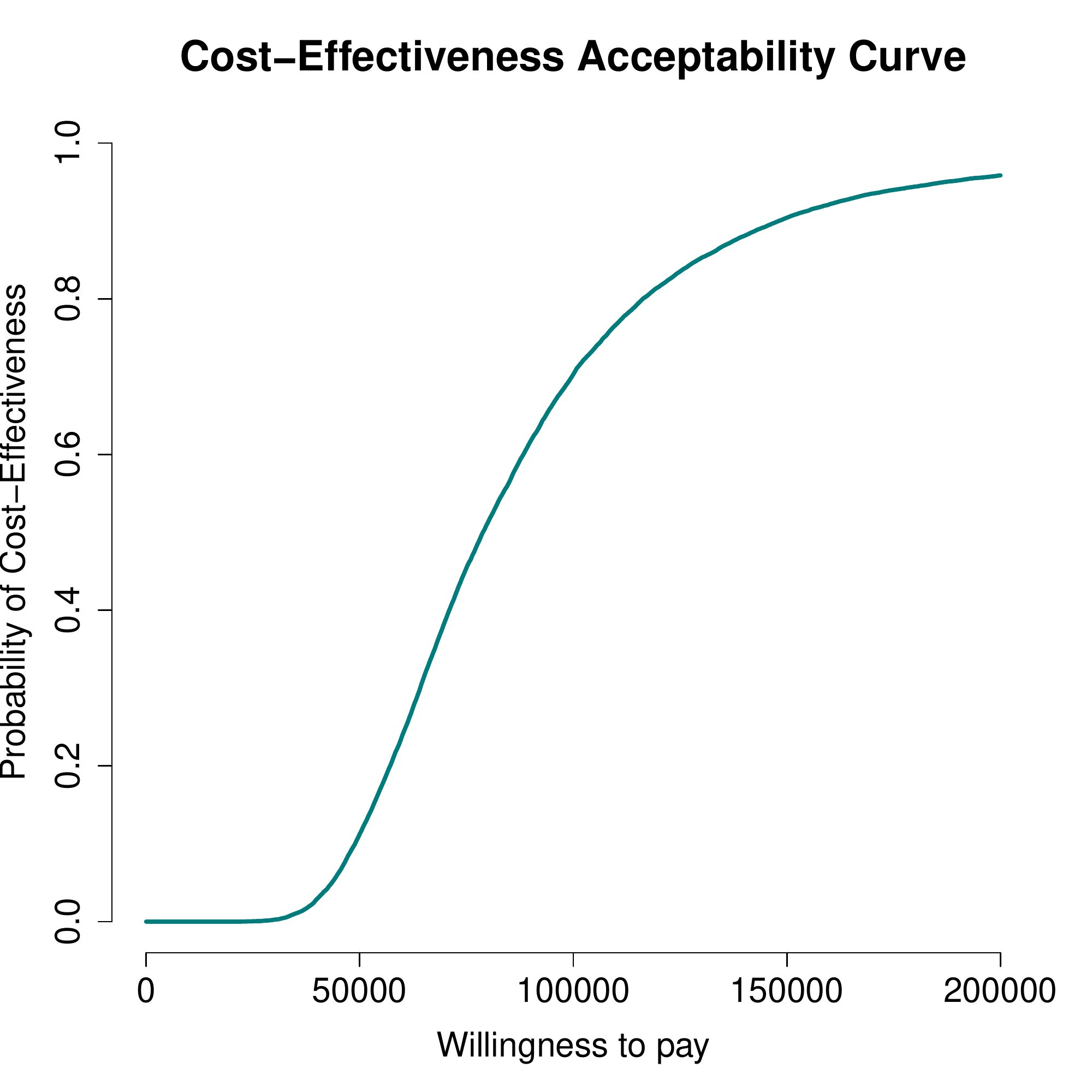}}
\caption{CEP (panel a) and CEAC (panel b) graphs associated with the two interventions in the TOPICAL trial. In the CEP, the value of the ICER is reported (darker green dot), while the portion of the plane on the right-hand side of the straight line passing through the origin (evaluated at $k=\pounds 55000$) denotes the sustainability area; in the CEAC, the probability of cost-effectiveness is shown for willingness to pay threshold values up to $\pounds 200000$.}\label{ee_plot}
\end{figure}

\end{document}